\documentclass[10pt,journal,compsoc]{IEEEtran}

\usepackage{ifpdf}
\usepackage{cite}
\usepackage{soul}
\usepackage[super]{nth}
\usepackage{lipsum}
\usepackage{xcolor}
\usepackage{listings}
\usepackage{calc}
\usepackage{wrapfig}
\newlength{\depthofsumsign}
\newcommand{\nsum}[1][1.4]{%
    \mathop{%
        \raisebox
            {-#1\depthofsumsign+1\depthofsumsign}
            {\scalebox
                {#1}
                {$\displaystyle\sum$}%
            }
    }
}

\colorlet{punct}{red!60!black}
\definecolor{background}{HTML}{EEEEEE}
\definecolor{darkgreen}{HTML}{006400}
\definecolor{darkorange}{HTML}{742902}
\definecolor{delim}{RGB}{20,105,176}
\colorlet{numb}{magenta!60!black}

\lstdefinelanguage{JavaScript}{
  morekeywords=[1]{break, continue, delete, else, for, function, if, in,
    new, return, this, typeof, var, void, while, with},
  morekeywords=[2]{false, null, true, boolean, number, undefined,
    Array, Boolean, Date, Math, Number, String, Object},
  morekeywords=[3]{eval, parseInt, parseFloat, escape, unescape},
  sensitive,
  morecomment=[s]{/*}{*/},
  morecomment=[l]//,
  morecomment=[s]{/**}{*/}, %
  morestring=[b]',
  morestring=[b]"
}[keywords, comments, strings]

\lstdefinelanguage{json}{
    basicstyle=\normalfont\ttfamily,
    numberstyle=\scriptsize,
    string=[s]{"\{}{\}"},
    stringstyle=\itshape\color{darkgreen},
    emph={microservices,hosts,routers,links,traffics,service_chains,traffic_types,traffic_type,nodes_settings,replica_count,res_scenario,placement_scenario,topology,nodes,edges,algorithm,options,anti-affinity,affinity}, 
    emphstyle=\color{blue},
    stepnumber=1,
    numbersep=8pt,
    showstringspaces=false,
    breaklines=true,
    frame=lines,
    backgroundcolor=\color{background},
    literate=
     *{0}{{{\color{numb}0}}}{1}
      {1}{{{\color{numb}1}}}{1}
      {2}{{{\color{numb}2}}}{1}
      {3}{{{\color{numb}3}}}{1}
      {4}{{{\color{numb}4}}}{1}
      {5}{{{\color{numb}5}}}{1}
      {6}{{{\color{numb}6}}}{1}
      {7}{{{\color{numb}7}}}{1}
      {8}{{{\color{numb}8}}}{1}
      {9}{{{\color{numb}9}}}{1}
      {:}{{{\color{punct}{:}}}}{1}
      {,}{{{\color{punct}{,}}}}{1}
      {\{}{{{\color{delim}{\{}}}}{1}
      {\}}{{{\color{delim}{\}}}}}{1}
      {[}{{{\color{delim}{[}}}}{1}
      {]}{{{\color{delim}{]}}}}{1},
}

\makeatletter
\def\lst@makecaption{%
  \def\@captype{table}%
  \@makecaption
}
\makeatother
\usepackage{actuarialsymbol}
\usepackage{caption}
\usepackage{subcaption}
\usepackage{siunitx}
\usepackage{enumitem}
\usepackage{ragged2e, eqparbox}
\usepackage{floatrow}
\usepackage{changepage}
\usepackage{makecell}
\usepackage{hhline}
\usepackage{pbox}
\usepackage{array}
\newcolumntype{P}[1]{>{\centering\arraybackslash}p{#1}}
\newcolumntype{M}[1]{>{\centering\arraybackslash}m{#1}}
\newlength\mytemplength

\newfloatcommand{capbtabbox}{table}[][\FBwidth]
\newcommand\parboxc[3]{%
    \settowidth{\mytemplength}{#3}%
    \parbox[#1][#2]{\mytemplength}{\centering #3}%
}

\ifCLASSINFOpdf
  \usepackage[pdftex]{graphicx}
  \graphicspath{{pdf/}{jpeg/}}
  \DeclareGraphicsExtensions{.pdf,.jpeg,.png}
\else
  \usepackage[dvips]{graphicx}
  \graphicspath{{../eps/}}
  \DeclareGraphicsExtensions{.eps}
\fi
\usepackage{algorithmic}
\usepackage[ruled,vlined, linesnumbered]{algorithm2e}
\usepackage{multicol}
\usepackage{amsmath,amssymb,amsfonts, mathtools, nccmath}

\usepackage{stfloats}
\ifCLASSOPTIONcaptionsoff
  \usepackage[nomarkers]{endfloat}
 \let\MYoriglatexcaption\caption
 \renewcommand{\caption}[2][\relax]{\MYoriglatexcaption[#2]{#2}}
\fi
\usepackage{url}
\makeatletter
\newsavebox\myboxA
\newsavebox\myboxB
\newlength\mylenA

\newcommand*\xoverline[2][0.75]{%
    \sbox{\myboxA}{$\m@th#2$}%
    \setbox\myboxB\null%
    \ht\myboxB=\ht\myboxA%
    \dp\myboxB=\dp\myboxA%
    \wd\myboxB=#1\wd\myboxA%
    \sbox\myboxB{$\m@th\overline{\copy\myboxB}$}%
    \setlength\mylenA{\the\wd\myboxA}%
    \addtolength\mylenA{-\the\wd\myboxB}%
    \ifdim\wd\myboxB<\wd\myboxA%
       \rlap{\hskip 0.5\mylenA\usebox\myboxB}{\usebox\myboxA}%
    \else
        \hskip -0.5\mylenA\rlap{\usebox\myboxA}{\hskip 0.5\mylenA\usebox\myboxB}%
    \fi}
\makeatother
\usepackage[pdf]{graphviz}
\usepackage{caption}
\usepackage{todonotes}
\usepackage{graphics}
\usepackage{physics}
\usepackage{tikz}
\usepackage{mathdots}
\usepackage{yhmath}
\usepackage{cancel}
\usepackage{color, colortbl}
\usepackage{siunitx}
\usepackage{multirow}
\usepackage{gensymb}
\usepackage{tabularx}
\usepackage{booktabs}
\usepackage{stmaryrd}
\usepackage{pifont}

\usetikzlibrary{fadings}
\usetikzlibrary{patterns}
\usetikzlibrary{shadows.blur}
\usetikzlibrary{shapes}
\usepackage{tikzscale}
\usepackage{filecontents}
\let\OrgPgfTransformScale\pgftransformscale
\renewcommand*{\pgftransformscale}[1]{%
  \gdef\ScaleFactor{#1}%
  \OrgPgfTransformScale{#1}%
}
\def\ScaleFactor{1}
\makeatletter
\newcommand{\shorteq}{%
  \settowidth{\@tempdima}{-}%
  \resizebox{\@tempdima}{\height}{=}%
}
\makeatother

\definecolor{code_indent}{HTML}{CCCCCC}

\newcommand{\indentrule}{\color{code_indent}\vrule\hspace{2pt}}

\begin{document}
\newcommand{\name}{PerfSim}

\title{\name: A Performance Simulator for Cloud Native Microservice Chains}

\author{Michel~Gokan Khan,~\IEEEmembership{Member,~IEEE,}
	Javid~Taheri,~\IEEEmembership{Senior~Member,~IEEE,}
	Auday~Al-Dulaimy,~\IEEEmembership{Member,~IEEE,}
	and~Andreas~Kassler,~\IEEEmembership{Senior~Member,~IEEE}	%
	
    \IEEEcompsocitemizethanks{\IEEEcompsocthanksitem M. Gokan Khan, J. Taheri, and A. Kassler are with the Department of Mathematics and Computer Science, Karlstad University, Sweden. \protect\\
    E-mails: \{michel.gokan,javid.taheri,andreas.kassler\}@kau.se.
    \IEEEcompsocthanksitem A. Al-Dulaimy is with  the School  of  Innovation,  Design  and  Engineering, Mälardalen University,  Sweden. E-mail: auday.aldulaimy@mdh.se}
}

\markboth{}%
{Shell \MakeLowercase{\textit{et al.}}: Bare Demo of IEEEtran.cls for IEEE Journals}

\IEEEtitleabstractindextext{%
\begin{abstract}

Cloud native computing paradigm allows microservice-based applications to take advantage of cloud infrastructure in a scalable, reusable, and interoperable way. However, in a cloud native system, the vast number of configuration parameters and highly granular resource allocation policies can significantly impact the performance and deployment cost. For understanding and analyzing these implications in an easy, quick, and cost-effective way, we present PerfSim, a discrete-event simulator for approximating and predicting the performance of cloud native service chains in user-defined scenarios. To this end, we proposed a systematic approach for modeling the performance of microservices endpoint functions by collecting and analyzing their performance and network traces. With a combination of the extracted models and user-defined scenarios, PerfSim can then simulate the performance behavior of all services over a given period and provide an approximation for system KPIs, such as requests' average response time. Using the processing power of a single laptop, we evaluated both simulation accuracy and speed of PerfSim in 104 prevalent scenarios and compared the simulation results with the identical deployment in a real Kubernetes cluster. We achieved {\raise.17ex\hbox{$\scriptstyle\mathtt{\sim}$}}81-99\% simulation accuracy in approximating the average response time of incoming requests and {\raise.17ex\hbox{$\scriptstyle\mathtt{\sim}$}}16-1200 times speed-up factor for the simulation.
	
\end{abstract}

\begin{IEEEkeywords}
	performance simulator, performance modeling, cloud native computing, service chains, simulation platform
\end{IEEEkeywords}
}
\IEEEpeerreviewmaketitle
\maketitle
\IEEEdisplaynontitleabstractindextext
\section{Introduction}
\IEEEPARstart{C}loud Native Computing is an emerging paradigm of distributed computing that ``\textit{empower organizations to build and run scalable applications in modern, dynamic environments such as public, private, and hybrid cloud}'' \cite{a2018_cncf}. One of the key purposes of introducing this paradigm was to answer the increasing need for mitigating the efforts of application-level clustering in the cloud and inline with the emergence of microservice architecture that promotes decoupling components of a software system into multiple independently manageable services, known as a microservice \cite{10.1145/3183628.3183631}. Since Google first introduced Kubernetes during the Google Developer Forum in 2014, as an approach for ``\textit{decoupling of application containers from the details of the systems on which they run}'' \cite{k8s_intro}, it becomes the de-facto enabler for utilizing microservice architecture based on technologies such as OS-level virtualization, better known as containers \cite{Casalicchio2019}.

Amongst the main advantages of cloud native computing is the possibility for allocating highly granular resources to large-scale chains of services in a cluster. This additional granularity, while facilitating the scalability of service chaining, imposes complexity in resource allocation, traffic shaping and placement of containers in a cluster. Therefore, as microservices networks grow larger, the need for tools and techniques that shed light on service chaining implications upon system performance becomes critical. Furthermore, by the rising trend of cloud native computing in containerized cluster environments, many researchers are nurturing new methods and schemes to implement new deployment \cite{10.1007/978-3-030-45989-5_6} and performance optimization techniques \cite{mao2020resource} at various levels: starting from modern container scheduling \cite{HU2020562,Liu2018ANC,Menouer2020KCSSKC,Mao2020SpeculativeCS} methods to predictive request load-balancing \cite{Burroughs2021} and resource auto-scaling algorithms \cite{Casalicchio2019-2,10.1007/978-981-13-9190-3_35}.

Analyzing the performance behavior of a service chain in a real testbed gives the most reliable results. However, evaluating in real clusters is not always possible. In many cases, it might be too costly, notably time-consuming, and sometimes various skill sets are required to configure, run and manage the testbed. Moreover, in most performance optimization techniques, various scenarios need to be evaluated in a timely manner to eventually minimize a cost function. Performing such evaluations in a real testbed, while providing accurate results, imposes a dramatic burden for achieving a scalable and efficient optimization method.

Moreover, modern cloud native distributed systems have important performance affecting properties that earlier generation software systems such as monoliths (single-tiered software systems consisting of multiple tightly-coupled components) didn't have to cope with as much. Properties such as agile horizontal/vertical scaling, highly granular resource allocation, context-awareness, contention with other services, service chaining, and dynamic load-balancing between replicas.

To mitigate the aforementioned implications of using real testbeds for performance evaluation of cloud native microservice chains, we proposed \textit{\name}, a simulation platform that aims to approximately predict the performance of service chains under various placement, resource allocation, and traffic scenarios using profoundly limited resources of a laptop. We also proposed a systematic performance modeling approach to model the time-predictable endpoint functions of microservices using performance traces generated by the profiling and tracing tools such as \textit{perf} and \textit{ebpf} as well as distributed network tracing programs such as \textit{jeager} and \textit{zipkin}. Using these models and a user-defined scenario, \name{} can then simulates performance behavior of all service chains under a desired placement, resource allocation policy, network topology and traffic scenario.

Using profoundly limited resources of a laptop, we evaluated the simulation accuracy and speed of PerfSim under 104 prevalent scenarios by deploying and running them on a real Kubernetes cluster and comparing measured KPIs with the simulation results (i.e., average requests latency). We used \textit{sfc-stress}, a synthetic service chain generation toolkit, for generating various service chains and microservice-based workloads. In our evaluation, we achieved {\raise.17ex\hbox{$\scriptstyle\mathtt{\sim}$}}81-99\% accuracy in predicting the average latency of incoming requests and {\raise.17ex\hbox{$\scriptstyle\mathtt{\sim}$}}16-1200 times speed-up factor between the simulation time and actual execution time on a real cluster. With the same laptop, we also simulated a large service chain consisting of 100 microservices interconnected with 200 links over 100 hosts and showed that \name{} can be effectively used for large-scale simulations.

To summarize, with \name{} we contributed to the relatively new, but rich area of cloud native computing by enabling a fast, accurate and easy way for evaluating various user-defined policies in microservice-based applications.

\section{Related Works}
In this section, we briefly review existing works in the areas of (1) simulation platforms, (2) emulation tools and (3) analytical performance modeling approaches. Table \ref{tab:comparison} presents a comparison between the key properties of the most popular frameworks with {\name}. In this table, besides common features, we also compared additional challenges imposed by simulating performance of microservices in cloud native environments.

\subsection{Simulation tools}
Cloud native systems have intricate provisioning and deployment requirements. Evaluating and predicting the performance of such services, studying the impact of provisioning policies, and correlating workload models with achievable performance are not straightforward due to the diversity and complexity of system interactions. Even though studying these interactions on real testbeds provides the most accurate results, several researchers and companies argue that computer simulation can be a powerful method to test multiple scenarios and evaluate various policies before enforcing them at different levels \cite{5683561,CloudSimToolkit2010,PULIAFITO2020102062,Riley2010}. 
In recent years, several works aim to predict the services' performance using computer simulations to develop adequate resource policies and other decisions at different levels to meet the required \textit{Quality of Service} (QoS) in a dynamic manner.

CloudSim \cite{CloudSimToolkit2010} is one of the most popular simulators in this category that is designed to simulate cloud computing infrastructures. It allows modeling of various data centers, workload scheduling, and allocation policies. Using CloudSim in many scenarios can boost the development of innovative methods and algorithms in the cloud computing paradigm without the need for deploying applications in the production environments. In recent years, several tools and modules had been designed based on CloudSim. For example, the iFogSim toolkit \cite{iFogSim2017} inherits the features of CloudSim, and extends them by the ability of modeling IoT and Edge/Fog environments. The authors in \cite{EdgeCoudSim2018} also built upon CloudSim to simulate the specifications of edge/fog computing and support the required functionalities. There exist several other CloudSim-based simulators for simulating various use cases in fog or edge environments \cite{7987304, RECAP2017-1}.

CloudSim and modules/plugins/tools based on CloudSim provide a sophisticated and straightforward way for researchers to simulate cloud/edge/fog computing infrastructure for modeling service brokers, data-centers, and scheduling policies. However, they are not designed with the purpose of simulating rigorous performance testings of microservices under extremely granular resource allocation and placement policies that exists in containerized services in cloud native applications. Moreover, to effectively simulate a set of service chains, we need to precisely specify the links and connections between each microservice within all chains, and then route the requests based on the user-defined traffic scenarios and network topology model. Something that the CloudSim category of simulators has not been designed to address. 

In addition to CloudSim, there are other simulators in the context of cloud simulation. \textit{Yet Another Fog Simulator} (YAFS) \cite{YAFS2019} is a discrete-event simulator based on \textit{Simpy} that allows to simulate the impact of applications' deployment in edge/fog computing environments through customizable strategies. YAFS allows to model the relationships between applications, infrastructure configurations, and network topologies. It uses those relationships to predict network throughput and latency in dynamic and customized scenarios, such as path routing and service scheduling. Even though YAFS offers a novel approach towards simulating performance of large network topologies, it's not designed to simulate the microservices' performance within the context of complex service chains over a set of hosts in a cluster.

Apart from cloud/edge/fog simulators, there exist other simulators that focus on specific aspects of cloud computing paradigm, such as energy efficiency or network modeling and optimization. For example, GreenCloud \cite{5683561}, is a packet-level simulator for capturing energy footprint of data center components with the aim towards providing an environment for researchers to design more energy-aware data centers \cite{7090996,6825028}. Other examples are NS-3 \cite{Riley2010}, OMNet++ \cite{Varga2019} and NetSim \cite{WAHIDULASHRAF2018547} that are primarily designed to simulate various types of networks and topologies. Although these simulators can smoothly simulate specific aspects of cloud systems that they are designed for (e.g., networking and energy efficiency), they cannot be used to simulate cloud native applications' performance; or at least can only be used to simulate specific aspects of service chains, such as their network performance or placement efficiency.

\subsection{Performance emulators}

Another popular experimentation approach in evaluating performance of cloud systems is emulation. Using emulation, users can analyze the system performance behavior supported by the available hardware in a more realistic way than using simulation \cite{beuran2012introduction}. Mininet \cite{MiniNetWebsite,10.1145/1868447.1868466} is one of the widely used network emulators that can form a network of virtual switches, routers, controllers, links, and hosts. The hosts in Mininet are able to run Linux network software, while switches offer flexible custom routing and SDN. The bright reputation of Mininet inspired many researchers to extend Mininet in various ways. For example, MaxiNet \cite{6857078} extended MiniNet to support spanning emulated networks over multiple hosts. Then EmuFog \cite{EmuFog2017} extended MaxiNet to support fog computing deployment models. Another MaxiNet-based emulator, called Fogbed \cite{Fogbed2018} enables the emulation of fog nodes as software containers under various network configurations to mimic edge/fog environments.

After Mininet introduced the idea of using containers for network and process emulation, other works also started to adopt a similar concept. For example, Dockemu \cite{to2015dockemu} adopted both Docker for emulating network nodes and NS-3 for simulating the network traffic. NEaaS \cite{lai2020network}, a cloud-based network emulation platform, utilizes both Docker and virtual machines to emulate various networking scenarios.

However, the widespread popularity of such emulators stands in stark contrast to the rather trembling fact that they are bounded to the available computation power and bandwidth of the underlying hardware they are deployed upon, and consequently, they cannot be efficiently used to predict the resource utilization aspects of large-scale and complex cloud native applications. Moreover, emulation cannot dramatically improve the speed of evaluating various resource allocation or placement policies, and therefore can't be used for exhaustive policy testing.

\subsection{Analytical performance modeling approaches}

\newcolumntype{G}{>{\columncolor{gray!10}}M}
\newcommand{\grey}{\cellcolor{gray!10}}
\newcommand{\nocolor}{\cellcolor{white}}

\begingroup
    \renewcommand*{\arraystretch}{1.5}%
    \definecolor{tabred}{RGB}{230,36,0}%
    \definecolor{tabgreen}{RGB}{0,116,21}%
    \definecolor{taborange}{RGB}{255,124,0}%
    \definecolor{tabbrown}{RGB}{171,70,0}%
    \definecolor{tabyellow}{RGB}{255,253,169}%
    \newcommand*{\redtriangle}{\textcolor{tabred}{\ding{115}}}%
    \newcommand*{\greenbullet}{\textcolor{tabgreen}{\ding{108}}}%
    \newcommand*{\orangecirc}{\textcolor{taborange}{\ding{109}}}%
    \newcommand*{\headformat}[1]{{\small#1}}%
    \newcommand*{\vcorr}{%
      \vadjust{\vspace{-\dp\csname @arstrutbox\endcsname}}%
      \global\let\vcorr\relax
    }%
    \newcommand*{\HeadAux}[1]{%
      \multicolumn{1}{@{}r@{}}{%
        \vcorr
        \sbox0{\headformat{\strut #1}}%
        \sbox2{\headformat{Complex Data Movement}}%
        \sbox4{\kern\tabcolsep\redtriangle\kern\tabcolsep}%
        \sbox6{\rotatebox{25}{\rule{0pt}{\dimexpr\ht0+\dp0\relax}}}%
        \sbox0{\raisebox{.5\dimexpr\dp0-\ht0\relax}[0pt][0pt]{\unhcopy0}}%
        \kern8mm %
        \rlap{%
          \raisebox{.25\wd4}{\rotatebox{25}{\unhcopy0}}%
        }%
        \kern.25\wd1 %
        \ifx\HeadLine Y%
          \dimen0=\dimexpr\wd2+.1\wd4\relax %
          \rlap{\rotatebox{25}{\hbox{\vrule width\dimen0 height .4pt}}}%
        \fi
      }%
    }%
    \newcommand*{\head}[1]{\HeadAux{\global\let\HeadLine=Y#1}}%
    \newcommand*{\headNoLine}[1]{\HeadAux{\global\let\HeadLine=N#1}}%
    \noindent
\begin{table*}[htbp]
	\caption{Comparison of performance characteristics in popular simulation and emulation frameworks with {\name}}
	\label{tab:comparison}
	\begin{footnotesize}
	\begin{flushleft}
	\resizebox{.85\columnwidth}{!}{%
    \begin{tabular}{%
      >{\bfseries}lc|>{\quad}c
      *{3}{c|}c>{\quad}c
      *{4}{c|}c>{\quad}c
      *{2}{c|}c>{\quad}c
      *{2}{c|}c%
    }%
      &
      \head{\footnotesize Type} &
      &
      \head{\footnotesize Target Environment} &
      \head{\scriptsize Granular Resource Allocation} &
      \head{\footnotesize Vertical Scaling} &
      \headNoLine{\footnotesize Horizontal Scaling} &
      &
      \head{\footnotesize CPU Scheduling} &
      \head{\scriptsize Multi-threaded Endpoints} &
      \head{\footnotesize Contention Modeling} &
      \head{\footnotesize Fast Performance Prediction} &
       \headNoLine{\footnotesize PTE Independent/Low Cost} &
      &
      \head{\footnotesize Service Chains Support} &
      \head{\footnotesize Multi-chain Support} &
      \headNoLine{\footnotesize Dynamic Control} &
      &
      \head{\footnotesize Load-balancing} &
      \head{\footnotesize Congestion Control} & \headNoLine{\footnotesize Advanced Routing}
      \\
      \sbox0{S}%
      \rule{0pt}{\dimexpr\ht0 + 2ex\relax}%
      CloudSim category \cite{CloudSimToolkit2010} & \textcolor{tabred}{\bfseries S} &&
      Cloud &  & \checkmark & &&
      \checkmark &  &  & \checkmark & \checkmark && &  & \checkmark &&
      \checkmark & \checkmark &  
      \\\hline
      YAFS \cite{YAFS2019} & \textcolor{tabred}{\bfseries S} &&
      Fog/Edge &  & & &&
       &  & & \checkmark & \checkmark &&
       \checkmark & \checkmark & \checkmark &&
       & \checkmark &  \checkmark
      \\\hline
      GreenCloud \cite{5683561} & \textcolor{tabred}{\bfseries S} &&
      Cloud &  & & &&
       &  & & \checkmark & \checkmark &&
       &  & \checkmark &&
      \checkmark &  &  
      \\\hline
      NS-3 \cite{Riley2010} & \textcolor{tabred}{\bfseries S} &&
      Any &  & & &&
       &  & & \checkmark & \checkmark &&
      \checkmark & \checkmark & \checkmark &&
      \checkmark & \checkmark & \checkmark 
      \\\hline
      OMNet++ \cite{Varga2019}  & \textcolor{tabred}{\bfseries S} &&
      Any &  & & &&
       &  & & \checkmark & \checkmark &&
       \checkmark & \checkmark & \checkmark &&
      \checkmark & \checkmark &  \checkmark
      \\\hline
      NetSim \cite{WAHIDULASHRAF2018547} & \textcolor{tabred}{\bfseries S} &&
      Any &  & & &&
       &  &  & \checkmark & \checkmark &&
      \checkmark & \checkmark & \checkmark &&
      \checkmark & \checkmark &  \checkmark
      \\\hline
      Emulators \cite{10.1145/1868447.1868466}\cite{to2015dockemu}\cite{lai2020network} & \textcolor{tabgreen}{\bfseries E} &&
      Any & \checkmark & \checkmark & \checkmark &&
      \checkmark & \checkmark & \checkmark & & &&
      \checkmark & \checkmark  & \checkmark  &&
       & \checkmark & \checkmark 
      \\\hline
      \rowcolor{tabyellow}%
      {\name} & \textcolor{tabred}{\bfseries S} &&
      Cloud/Edge & \checkmark & \checkmark & \checkmark &&
      \checkmark & \checkmark & \checkmark & \checkmark & \checkmark &&
      \checkmark & \checkmark & \checkmark &&
      \checkmark & \checkmark &  
      \\[.5ex]
      \multicolumn{4}{c}{\scriptsize \textcolor{tabred}{\bfseries S}=Simulator,\textcolor{tabgreen}{\bfseries E}=Emulator
      } &
      \multicolumn{3}{c}{\bfseries Resources} &&
      \multicolumn{4}{c}{\bfseries Threads} &&
      \multicolumn{3}{c}{\bfseries Services}
      &&
      \multicolumn{3}{c}{\bfseries Network}
      \\
    \end{tabular}%
	}
		\end{flushleft}
	\end{footnotesize}
    \end{table*}
    \kern19.5mm %
  \endgroup

There is a third category of approaches towards modeling the performance of cloud native applications which based upon an analytical framework. Contrary to the simulation/emulation techniques, which employ a bottom-up approach in modeling and simulating various performance aspects of cloud applications, analytical methods adopt a top-down approach for that purpose by collecting and analyzing application KPIs via running various stress tests on the system or using historical performance measurements.

As an example, the work proposed in \cite{LearningPredictiveAutoscaling2019} aims to model the response time of microservices based on stress testing and concurrently collecting performance traces for predefined intervals to learn a predictive auto-scaling model, such that the response time requirements are satisfied. Those performance traces are used to learn the resource provisioning policy model using a regression analysis approach. In \cite{PerformanceModelingfor2019}, microservices are being tested individually using service-based sandboxing to construct the corresponding model, the analyzed data and performance model are presented to the user. 
Considering modeling throughput and latency, in our previous work \cite{APerformanceModellingApproach2020}, we used a combination of stress-testing and regression modeling approaches to understand the impact of microservices’ resource configurations, model the correlation between the KPIs and resource configurations in a smaller \textit{Performance Testing Environment} (PTE) to predict the performance in a larger \textit{Production Environment} (PE).

Although the aforementioned methods benefit from accurate performance measurements on real systems and can be considered as the most widely used techniques in performance engineering in production environments, they suffer from three major limitations. Their first limitation is the high cost of test environment preparation that can accurately provide performance measurements. As for gathering accurate performance insights, the tests need to be performed on either a production environment or a PTE that mimics it; that imposes a high cost of deployment in either case. The second limitation is the slow stress-testing procedure as for getting accurate results, several hours of performance testing is required. Last but not least of their limitations is applicability in utilizing advanced optimization techniques, such as deep learning or meta-heuristics based optimization heuristics, which requires numerous re-deployments to learn, train, or find an optimal policy.

\section{{\name} Design and Implementation}
In this section, we introduce the system architecture and implementation details of {\name} and mathematical notation used in this paper (summarized in Table \ref{tab:notations}).

\subsection{Modeling elements of cloud native systems}

{\name} has models for various elements of cloud native systems and their underlying infrastructure.

As described in the previous sections, the joint problem of placement and resource allocation of a service chain over a cluster is complex as multiple dimensions impact the achievable performance and latency.  However, several parameters have only marginal impact on the prediction quality. Consequently, we aim at providing a simplified model that is tractable but powerful enough in order to predict the performance of typical cloud native service chain architectures when deployed on several well-known container orchestration platforms such as Kubernetes.

\begin{table*}[htbp]
	\small
	\setlength\tabcolsep{5.05pt}
	\setlength\aboverulesep{0pt}
	\setlength\belowrulesep{0pt}
	\setlength\doublerulesep{2pt}
	\caption{Table of the base notations used in this paper}
	\label{tab:notations}
	\resizebox{.96\columnwidth}{!}{%

	\begin{tabular*}{1.2835\linewidth}{|M{19pt}|p{68pt}|p{105pt}||p{77pt}|p{96pt}||p{76pt}|p{145pt}|}
		\toprule
		\Xhline{4\arrayrulewidth}
		 \multirow{2}{*}{\rotatebox[origin=c]{90}{\textbf{Hosts} }}&\parboxc{c}{22pt}{\(H=\{h_k\}_{k=1}^{|H|}\)} & Set of hosts &
		\parboxc{c}{22pt}{\(W^{r^H}\)} & Weights of resources & \parboxc{c}{22pt}{\(h_k^{\text{cores}},h_k^{\text{clock}}\)} & Cores count and CPU clock \footnotesize(hz) in \(h_k\) \\
		\cline{2-7}
		&\parboxc{c}{22pt}{\(r^H\in{R^H}\)} & Set of resource names &
		\parboxc{c}{22pt}{\(\Hat{V}_H(r^H,h_k)\)} & \footnotesize Initial \(r^h\) capacity of \(h_k\) &
		\parboxc{c}{22pt}{\({V_H(r^H,h_k)}\)} & \footnotesize Current \(r^H\) capacity of \(h_k\) \\
		\hline
		\hline
		\Xhline{4\arrayrulewidth}

		\multirow{3}{*}{\rotatebox[origin=c]{90}{\parbox{53pt}{\centering\footnotesize{\textbf{Network\\Topology}}}}}&\parboxc{c}{22pt}{\(P=\{\rho_z\}_{z=1}^{|P|}\)} & Set of all routers &
		 \parboxc{c}{22pt}{\(L=~\!\!^{\scalebox{.6}{\(H\)}}\!{L}\cup{^{\scalebox{.7}{\(\tau\)}}\!{L}}\)} & \parboxc{c}{22pt}{Set of all network links} & \(G(\tau)=(P,L)\) & \footnotesize Graph represent. of network topology \(\tau\)
 		 \\
		\cline{2-7}
		&\parboxc{c}{22pt}{\(^{\scalebox{.6}{\(H\)}}\!{L}=\{^{\scalebox{.6}{\(H\)}}\!{l}_{o}\}_{o=1}^{|^{\scalebox{.5}{\(H\)}}\!{L}|}\)} & \parboxc{c}{22pt}{Set of host\(\leftrightarrows\)router links} &
		\(\rho_z^{\text{in bw}},\rho_z^{\text{out bw}}\) & In/out bw of \(\rho_z\) (Bps) & \parboxc{c}{22pt}{\(G(\tau)_{h_a,h_b}\)} & \parboxc{c}{22pt}{\footnotesize Network path between \(h_a\) and \(h_b\)}
		\\
		\cline{2-7}
		& 	 \parboxc{c}{22pt}{\(^{\scalebox{.7}{\(\tau\)}}\!{L}=\{^{\scalebox{.7}{\(\tau\)}}\!{l}_o\}_{o=1}^{|^{\scalebox{.7}{\(\tau\)}}\!{L}|}\)} & \parboxc{c}{22pt}{Set of router\(\leftrightarrows\)router links}  &
		\(l_o^{\text{lat}},\rho_z^{\text{lat}}\) & Latencies of \(l_o\) \!\scriptsize\&\footnotesize\! \(\rho_z\) (ns) &
		\(\rho^{h_k}\in{P}\) & \footnotesize The router connected to \(h_k\)\\
		\hline
		\hline

		\Xhline{4\arrayrulewidth}

		\multirow{3}{*}{\rotatebox[origin=c]{90}{\parbox{56pt}{\centering\footnotesize{\textbf{Services and\\ Endpoints}}}}}&\parboxc{c}{22pt}{\(S=\{S_i\}_{i=1}^{|S|}\)} & Set of all services &
		\(r^S\in{R^S}\) & Set \footnotesize of \normalsize res. ctrl. params &
		\(\Pi=(p_{i,j}\in{\{0,1\}})\) & Service replica placement matrix \\
		\cline{2-7}
		&\parboxc{c}{22pt}{\(S_i=\{\Hat{s}_j^i\}_{j=1}^{|S_i|}\)} & Set of replicas of \(S_i\) &
	\parboxc{c}{22pt}{\(\Hat{V}_S(r^S\!,S_i)\),\(V_S(r^S\!,\hat{s}_j^i)\)} & \footnotesize Initial/current \(r^S\) \!cap. of \(\hat{s}_j^i\) &
		 \(\Pi(\Hat{s}_j^i)\in{H}\) & The host that \(\Hat{s}_j^i\) is currently placed \\
		\cline{2-7}
		&\parboxc{c}{22pt}{\(F_i=\{f_n^i\}_{n=1}^{|F_i|}\)} & \footnotesize Set  of endpnt. funcs of \(S_i\) &
		\(A_{S_i},\widetilde{A}_{S_i}\) & \footnotesize Affinities\tiny/\footnotesize anti-affinities &
		\(H^{\Hat{s}_j^i}\), \(\psi_{\Hat{s}_j^i}^{h_k}\) & Eligible hosts and score of \(h_k\) for \(\Hat{s}_j^i\) \\  \hline
		\hline
		\Xhline{4\arrayrulewidth}

		\multirow{3}{*}{\rotatebox[origin=c]{90}{\parbox{55pt}{\centering\footnotesize{\textbf{Service Chains\\and Traffic}}}}}&\parboxc{c}{22pt}{\(C=\{C_l\}_{l=1}^{|C|}\)} & \parboxc{c}{22pt}{Set of service chains} &
		\(\!\!~^{l}\!F\)=\(\{\!\!~^{l}\!f_{n}\}_{n=1}^{|\!\!~^{l}\!F|}\) & \parbox[c][22pt]{120pt}{\footnotesize Set of endpnt. funcs in \(C_l\)} &\(\!\!~^{l}\!U\)=\(\{\!\!~^{l}\!u_{o}\}_{o=1}^{|\!\!~^{l}\!U|}\) & \footnotesize Set of user requests in \(C_l\) 
		\\
		\cline{2-7}
		&\parboxc{c}{22pt}{\(\!\!~^{l}\!S=\{\!\!~^{l}\!S_{i}\}_{i=1}^{|\!\!~^{l}\!S|}\)\(\subseteq{S}\)} & Set of services in \(C_l\) &
		\(\!\!~^{l}\!F_i\)=\(\{\!\!~^{l}\!f^{i}_{n}\}_{n=1}^{|\!\!~^{l}\!F_i|}\!\subseteq{\!\!\!~^{l}\!F}\) & \footnotesize Set of endpnt. funcs of \(\!~^{l}\!S_{i}\) &
		\parboxc{c}{22pt}{\(C_l^{\text{rate}},\!C_l^{\text{duration}},\!C_l^{\text{batch}}\)} & \footnotesize Arrival rate/duration \& batch size of \(C_l\)\\ 
		\cline{2-7}
		&\parboxc{c}{22pt}{\(\!\!~^{l}\!E=\left(\!\!~^{l}\!e_{x}\right)_{x=1}^{|\!\!~^{l}\!\!E|}\)} & \parboxc{c}{22pt}{\footnotesize Ordered set of links in \(C_l\)} &
		\(G(C_l)\)=\((\!~^{l}\!F\),\(~^{l}\!E)\) & Graph represent. of \(C_l\) &
		\footnotesize \parboxc{c}{22pt}{\(\!\!~^{l}\!u_{o}^{\text{in time}},\!\!~^{l}\!u_{o}^{\text{exe time}}\)} & \parboxc{c}{22pt}{\footnotesize Req. arrival \& exe time of \(\!\!~^{l}\!u_o\)} \\ 
		
		\hline
		\hline
		\Xhline{4\arrayrulewidth}

        \multirow{3}{*}{\rotatebox[origin=c]{90}{\parbox{55pt}{\centering\footnotesize{\textbf{Threads}}}}}&\parboxc{c}{22pt}{\(\!\!~^{l}\!f_{n}\)=\(\{t_m\}_{m=1}^{|\!\!~^{l}\!f_{n}|}\)} & Set of threads in \(\!\!~^{l}\!f_n\) 
		&
		\parboxc{c}{22pt}{\(t_m^{\text{maccs}}\)} & Mem. accesses of \(t_m\) & \parboxc{c}{22pt}{\(t_m^{\text{blk rw}}\)} &  \parboxc{c}{22pt}{blkio R/W of \(t_m\) (in bytes)}
		\\
		\cline{2-7}
		&\parboxc{c}{22pt}{\(\!\!~^{l}\!f^{i}_{n}\)=\(\{t_m|t_m\in{\!\!~^{l}\!f^{i}_{n}}\}\)} & Set of threads in \(\!\!~^{l}\!f_n^i\subseteq{\!\!~^{l}\!f_{n}}\) &
		\parboxc{c}{22pt}{\(t_m^{\text{crefs}},t_m^{\text{cmiss}}\)} & \parboxc{c}{22pt}{\footnotesize{Cache refs \& misses of \(t_m\)}} &\(t_m^{\text{idle}}\)
		& Total idle time of \(t_m\)
		\\
		\cline{2-7}
		&\parboxc{c}{22pt}{\(t_m^{\text{inst}},t_m^{\text{CPI}}\)} & \parboxc{c}{22pt}{Inst. count \& CPI \footnotesize of \normalsize \(t_m\)} & \parboxc{c}{22pt}{\(t_m^{\text{cpenalty}}\)} &\footnotesize Avg. lost cycles per miss& \(\Hat{f}_{h_k}^{}\) & \footnotesize Set of running threads on \(h_k\) \\
		\hline

		\bottomrule
	\end{tabular*}

	}
	
\end{table*}
\subsubsection{Hosts}

An \(h_k\in{H=\{h_k\}_{k=1}^{|H|}}\) is a model of Linux-based physical machine that has \(h_k^{\text{cores}}\) number of \textit{CPU cores} with \(h_k^{\text{clock}}\) \textit{clock speed} (in Hertz). It has also a set of consumable resources, such as a \textit{memory capacity} (in bytes), a NIC with a limited \textit{ingress} and \textit{egress} \textit{network bandwidth} (in bytes per second), a local storage with a limited \textit{storage read/write bandwidth} and \textit{storage capacity}. The CPU resource is measured in \textit{millicores}. Each host \(h_k\) introspects the OS to determine the \(h_k^{\text{cores}}\) and then multiples it by $1000$ to denote its total capacity (Eq. \ref{millicores}).

\begin{equation}
\label{millicores}
    h_k^{\text{millicores}}=h_k^{\text{cores}}\times{1000}
\end{equation} 

To facilitate our formulation, we denote hosts consumable resources as \(R^H=\{\)millicores, mem, in bw, out bw, blkio bw, blkio size\(\}\). For each resource \(r^H\in{R^H}\), a host \(h_k\) has an initial resource capacity denoted as \(\Hat{V}_H(r^H,h_k)\) as well as a current capacity denoted as \(V_H(r^H, h_k)\).

\subsubsection{Network topology}
A network topology, denoted as \(\tau\), has a directed acyclic graph representation of \(G(\tau)=(P, ^{\scalebox{.7}{\(\tau\)}}\!\!{L})\) with routers \(\rho_z\in{P}\) as nodes, and directed links \(^{\scalebox{.7}{\(\tau\)}}\!{l}_o=(\rho_i,\rho_j)\in{^{\scalebox{.7}{\(\tau\)}}\!{L}}\) between them as edges. To separately model egress and ingress bandwidth of interconnection between nodes, we hypothetically assumed there are always 2 links between each pair of routers with opposite directions. We denote the maximum ingress/egress bandwidth of a router as \(\rho_z^{\text{in bw}}\) and \(\rho_z^{\text{out bw}}\). 

Each host \(h_k\) is connected to a router \({\rho}^{h_k}\in{P=\{\rho_z\}_{z=1}^{|P|}}\) using a hypothetically separated egress/ingress directed links, denoted as \(^{\scalebox{.6}{\(H\)}}\!{l}_o\in{^{\scalebox{.6}{\(H\)}}\!{L}}\) that have a pair of ordered host\(\rightarrow\)router \((h_k,\rho_i)\) and router\(\rightarrow\)host \((\rho_i,h_k)\)) links.

Each router \(\rho_z\) and link \(l_o\in{L}\), may respectively add extra latency of \(\rho_z^{\text{lat}}\) and \(l_o^{\text{lat}}\) nanoseconds to each request. These additional latencies differs from the delay caused by network congestion which we modelled separately.

\subsubsection{Services} 
We assume that each chain of services 
is composed of a set of \(|S|\) containerized \textit{services} \(S=\{S_i\}_{i=1}^{|S|}\) such that each \(S_i\in{S}\) may have \(|S_i|\) number of single-process/multi-threaded \textit{replicas} \(\{\Hat{s}_{j}^{i}\}_{j=1}^{|S_i|}\) load balanced among \(|H|\) \textit{hosts} and connected through \(|P|\) routers. A given service \(S_i\) has a set of endpoint functions \(F_i=\{f_n^i\}_{n=1}^{|F_i|}\) that can be executed based on the type of incoming request. Each endpoint function \(f_n^i\) may spawns a set of threads \(t_m\in{f_n^i}\). We will explain properties of these threads in section 3.1.10.

\subsubsection{CPU scheduler and resource controller}
A service \(S_i\) may optionally have resource constraints on its containerized replicas. We denote this set of resources as \(R^S=\{\)CPU requests, mem requests, in bw, out bw, blkio bw, blkio size\(\}\), which corresponds to resources \(R^H\) in each host (\(R:R^S\rightarrow{R^H}\)). By default, there is no reservation or usage limitation on any resource.

In Linux kernel, a service replica is in a form of a container and is being managed by \textit{Control Groups} (cgroups) \cite{cgroups} that is responsible for auditing and restricting a set of processes.  The limits associated with cgroups isolate the resource usage of a collection of processes. 

We partially modeled the behaviour of cgroups in \name{} by allowing a service \(S_i\) to initially reserve \(\Hat{V}_S(r^S, S_i)\) resource capacities for each of its replicas.  When reserving CPU requests \(\Hat{V}_S(\text{CPU requests},S_i)\), PerfSim allows a replica to use as much as CPU units available, even more than its reserved quota. We also partially modeled the behaviour of CPU bandwidth controller \cite{36669} of Linux by allowing to define \(S_i^{\text{CPU limits}}\) (in millicores) as an upper bound for replicas' CPU consumption. In kernel level, CPU request and limit is translated to cgroups \texttt{cpu.shares}, \texttt{cpu.cfs\_quota\_us} (in microseconds) and \texttt{cpu.cfs\_period\_us} (in microseconds) control parameters. We denote them as \(S_i^{\text{CPU share}}, S_i^{\text{CPU quota}}\), and \(S_i^{\text{CPU period}}\). However, for simplicity, we assume a fixed value for \(S_i^{\text{CPU period}}\)=100ms in our calculations and only considered CPU share and quota when estimating available resources for each replica. A usage example for this model is simulating Kubernetes \textit{Best Effort} and \textit{Guaranteed} QoS classes for CPU resources which we will cover in section 5.2.

The \(\Hat{V}_S(\text{mem requests},S_i)\) specifies the initial memory allocation of replicas of \(S_i\) (in bytes). We modelled the \texttt{memory.limit\_in\_bytes} control parameter in cgroups in a way that it may only affect the placement of replicas.

To control the traffic shaping and the maximum network bandwidth of replicas in \(S_i\), the \(\Hat{V}_S(\text{in bw},S_i)\), and \(\Hat{V}_S(\text{out bw},S_i)\) control parameters can be used (in Bps).

The storage size of replicas in \name{} is controlled by a model of \textit{blkio} cgroups controller and denoted as \(\Hat{V}_S(\text{blkio size},S_i)\). We also denote the available bandwidth for blkio reads and write as \(\Hat{V}_S(\text{blkio bw},S_i)\). We intended to model and simulate the disk bandwidth throttling feature of cgroups \texttt{blkio.throttle.write\_bps\_device}.

\subsubsection{Affinity controller}
A service may optionally have a set of in-service affinity/anti-affinity rules for its placement. 
We define two binary decision vectors \(A_{S_i}\) and \(\widetilde{A}_{S_i}\) that are respectively holding all the affinity and anti-affinity rules related to service \(S_i\). Each vector has a dimension of \(|S| \times 1\) where each row represents one of the services. An entry \(A_{S_i}[S_j]\) takes on the value of `1' if replicas of services \(S_i\) and \(S_j\) must be placed at the same host and `0' otherwise; and similarly an entry \(\widetilde{A}_{S_i}[S_j]\) takes on a value of `1' if any replicas of both services must not be placed at the same host and `0' otherwise. It's also possible to define host affinities, but due to the page limitation we omitted the formal definition.

\subsubsection{Placement algorithm}
The placement of each \(\Hat{s}_{j}^{i}\) among all hosts, takes place using a placement algorithm. By default, we implemented the \textit{Least Allocated} bin packing algorithm which partially simulates the container scheduling mechanism of Kubernetes. This algorithm, favor hosts with fewer resource requests and is controlled with weights for each resource \(r^H\in{R^H}\), denoted as \(W^{r^H}\) to be used for scoring nodes. We described the details of this algorithm in the next section.

\subsubsection{Container scheduler}
We use a binary matrix \(\Pi\) of size \((\sum_{i=1}^{|S|}{|S_i|})\times{|H|}\) to  keep tack of service replicas placement. In this matrix, \(p_{j,k}\in{\Pi}=1\) if the replica \(\Hat{s}_j\in\{S_i\}_{\in{S}}\) is placed in host \(h_k\) and \(p_{j,k}\)=\(0\) if otherwise. Since each replica can only be placed on one host, \(\Pi(\Hat{s}_i^j)\in{H}\) indicates the host that replica \(\Hat{s}_i^j\) is placed. After placing each replica, the amount of available resources in each host \(V_H(r^H,h_k)\) are recalculated for all resources.

\subsubsection{Service chains}
The goal of services is to serve requesters through a set of service chains \(C=\{C_l\}_{l=1}^{|C|}\). These service chains represent the connection and traffic flow between endpoint functions of services and are specified as a directed flow network \(G(C_l)\)=\((\!\!~^{l}\!F,\!\!~^{l}\!E)\) where \(\!\!~^{l}\!F\) is the subset of endpoint functions which have a role in providing service chain \(C_l\), and \(\!\!~^{l}\!E\) is the subset of all virtual links belonging to the service chain \(C_l\). In other words, all endpoint functions within \(C_l\) are connected through \(|\!\!~^{l}\!E|\) number of ordered directed virtual links \(\!\!~^{l}\!E\)=\(\big(\!~^{l}\!e_{1}\)=\((\!\!~^{l}\!f_a,\!\!~^{l}\!f_{a'}\!),...,\!\!~^{l}\!e_{|\!\!~^{l}\!E|}\)=\((\!\!~^{l}\!f_c,\!\!~^{l}\!f_{c'})\big)\). Each \(\!\!~^{l}\!e_{v}\in{\!\!~^{l}\!E}\) has a request payload \(\!\!~^{l}\!e_{v}^\text{payload}\) in bytes, which indicates the request size flowing from edge's first vertex to the second. 

The first node in a \(C_l\) is its only request entry point, known as the \textit{source} node, that receives requests from potential request senders and the cluster load-balancer directs the traffic to a replica \(\Hat{s}_j^i\) based on a desired load-balancing algorithm (default is \textit{Round-robin}). A service chain has also one or more \textit{sink} nodes with no immediate outgoing edges.

\subsubsection{Traffic control}
Assuming a service chain \(C_l\) is active for \(C_l^{\text{duration}}\) seconds. The traffic rate of each \(C_l\) is being controlled with 2 control parameter \(C_l^{\text{rate}}\) and \(C_l^{\text{batch}}\); every \(\frac{1}{C_l^{\text{rate}}}\) seconds, a batch of \(C_l^{\text{batch}}\) requests will arrive to service chain \(C_l\). In other words, each service chain \(C_l\) may be requested by a plurality of \(C_l^{\text{batch}}\) users with an average arrival rate of \(C_l^{\text{rate}} \frac{\text{req}}{\text{s}}\). Therefore, the number of requests \(|\!\!~^{l}\!U|\) for service chain \(C_l\) over a period of time \(C_l^{\text{duration}}\) will be equal to \(C_l^{\text{duration}}\times{C_l^\text{rate}}\times{C_l^{\text{batch}}}\). We denote set of all user requests of \(C_l\) as \(\!\!~^{l}\!U=\{\!\!~^{l}\!u_{o}\}_{o=1}^{|\!\!~^{l}\!U|}\) and their arrival time and execution time are denoted by \(\!\!~^{l}\!u_{o}^{\text{in time}}\) and \(\!\!~^{l}\!u_{o}^{\text{exe time}}\), respectively.

\subsubsection{Service endpoint functions and associated threads}
During the life cycle of the \(C_l\), each endpoint function \(\!\!~^{l}\!f_{n}\in{\!\!~^{l}F}\) spawns \(|\!\!~^{l}\!f_{n}|\) number of threads \(\{t_m\}_{m=1}^{|\!\!~^{l}\!f_{n}|}\) on one of its replicas. Since each replica is placed on a host, we denote a subset of all threads running on a host \(h_k\) as \(\Hat{f}_{h_k}^{}\). 

Each thread \(t_m\in{\!\!~^{l}\!f_{n}}\) has a set of properties that specified in the original model and extracted based on measurements of the performance traces and monitoring tools during the modeling phase which we will explain in section \ref{modelin_process_section}. 

A \(t_m\) executes \(t_m^{\text{inst}}\) CPU instructions in which \(t_m^{\text{maccs}}\) number of them are due to memory accesses with \(t_m^{\text{CPI}}\) average CPU \textit{Cycle per Instruction} (CPI). Moreover, \(t_m\) has total of \(t_m^{\text{crefs}}\) cache references in which, when \(\Hat{s}_j^i\) is deployed on an isolated and single core reference machine, \(t_{m}^{\text{cmiss}}\) number of those references will be cache misses with average miss penalty of \(t_{m}^{\text{cpenalty}}\) cycles. During the simulation, \name{} dynamically re-calculates cache misses based on various factors such as co-located active threads, number of cores, CPU requests, and cache size.

A thread may also read/write a total of \(t_m^{\text{blk rw}}\) bytes of data from/to the storage device. Moreover, a thread might be idle for some time during its active life-cycle and we denote \(t_m\) accumulated period of the idle time as \(t_m^{\text{idle}}\).

On a real Linux machine, threads are load balanced among all CPU runqueues in the host using the Linux's \textit{Completely Fair Scheduler} (CFS). We implemented a simplified version of CFS in \name{} to imitate the impact of thread scheduling mechanism on the performance. We described the details of our implementation in section 5.

\subsection{Modeling process}
\label{modelin_process_section}
\begin{figure*}[t]
	\centering
	\includegraphics[width=0.8\linewidth]{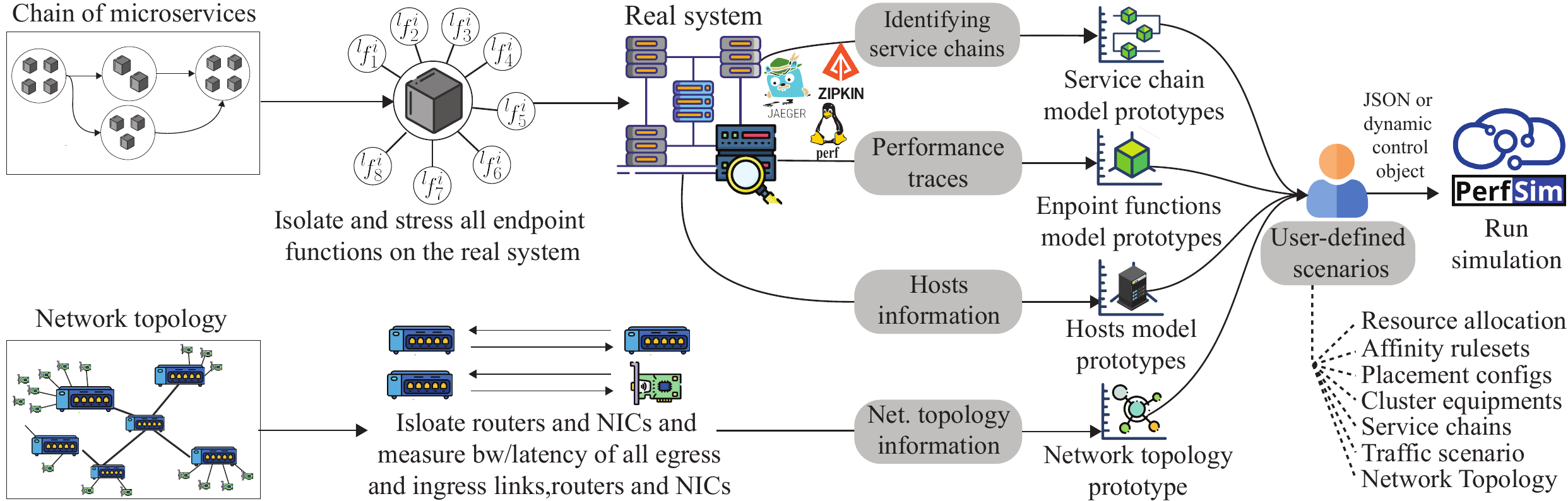}
	\caption{An overview of our proposed systematic method for modeling various service chains and microservices with their corresponding endpoint functions as well as underlying network topologies encapsulated in user-defined scenarios}
	\label{fig:modeling-process}
\end{figure*}

In order to accurately simulate the performance of service chains in a given scenario, PerfSim requires to have performance models of each and every active endpoint function within running services in the chain. To introduce such performance models in \name{}, the user can either provide a pre-defined model file, or run a new modelling procedure. In the latter case, we proposed a systematic performance testing and modeling approach to extract the KPIs of endpoint functions, construct aforementioned modeling elements, and identify each equipment's resource capacities in the simulation. Once such performance models being extracted for an endpoint function, the models can be re-used for simulating various types of user-defined scenarios without any need for a new profiling phase.

As presented in Figure \ref{fig:modeling-process}, the modeling process starts by placing microservices of all service chains on a set of reference hosts in the \textit{Performance Testing Environment} (PTE). Then, one by one the microservices get isolated on a single host to start stressing each of its endpoint functions by flowing a single request through the function, and in parallel, an automated script captures various performance traces and resource utilization measurements and stores the correlated traces it in a separate database. 

While running the network traffic, various network tracing tools would be used to identify the connections between microservices and identify service chains within the cloud native system. For example, a \textit{service mesh} layer based on \textit{istio} \cite{istio}, enables distributed tracing capabilities through Envoy \textit{\cite{envoy}} which ``\textit{allows developers to obtain visualizations of call flows in large service oriented architectures}'' by taking advantage of tracing tools such as \textit{LightStep}\cite{lightstep}, \textit{Zipkin}\cite{zipkin}, and \textit{Jaeger} \cite{jaeger}. Leveraging a service mesh layer allows \name{} to extract service chain flow models which facilitates the modeling procedure. However, these models can also be provided manually via \name{}'s JSON model files or dynamic control objects.

After completing the profiling phase for all endpoint functions, a similar procedure can optionally be repeated with all the target network equipments, such as routers and NICs, using network performance measurement tools such as \textit{iperf}. The intention is to separately measure available egress/ingress bandwidths of each network equipment and links latency to effectively simulate the network congestion.

Finally, based on these extracted models, various user-defined scenarios can be specified by the user. These scenarios includes resource allocation policies, affinity rulesets, placemnent conigurations and algorithm, equipments used in a cluster (i.e., active hosts and router), service chains, incoming traffic scenarios, and underlying network topology. To start simulating the given scenario, all the aforementioned models and simulation parameters can be fed into \name{} using a JSON model file or a Python script (i.e. dynamic control object). We will cover the details of defining a scenario in section 3.3.1.

\subsection{System Architecture}

\begin{figure}[hbtp]
	\centering
	\includegraphics[width=0.85\linewidth]{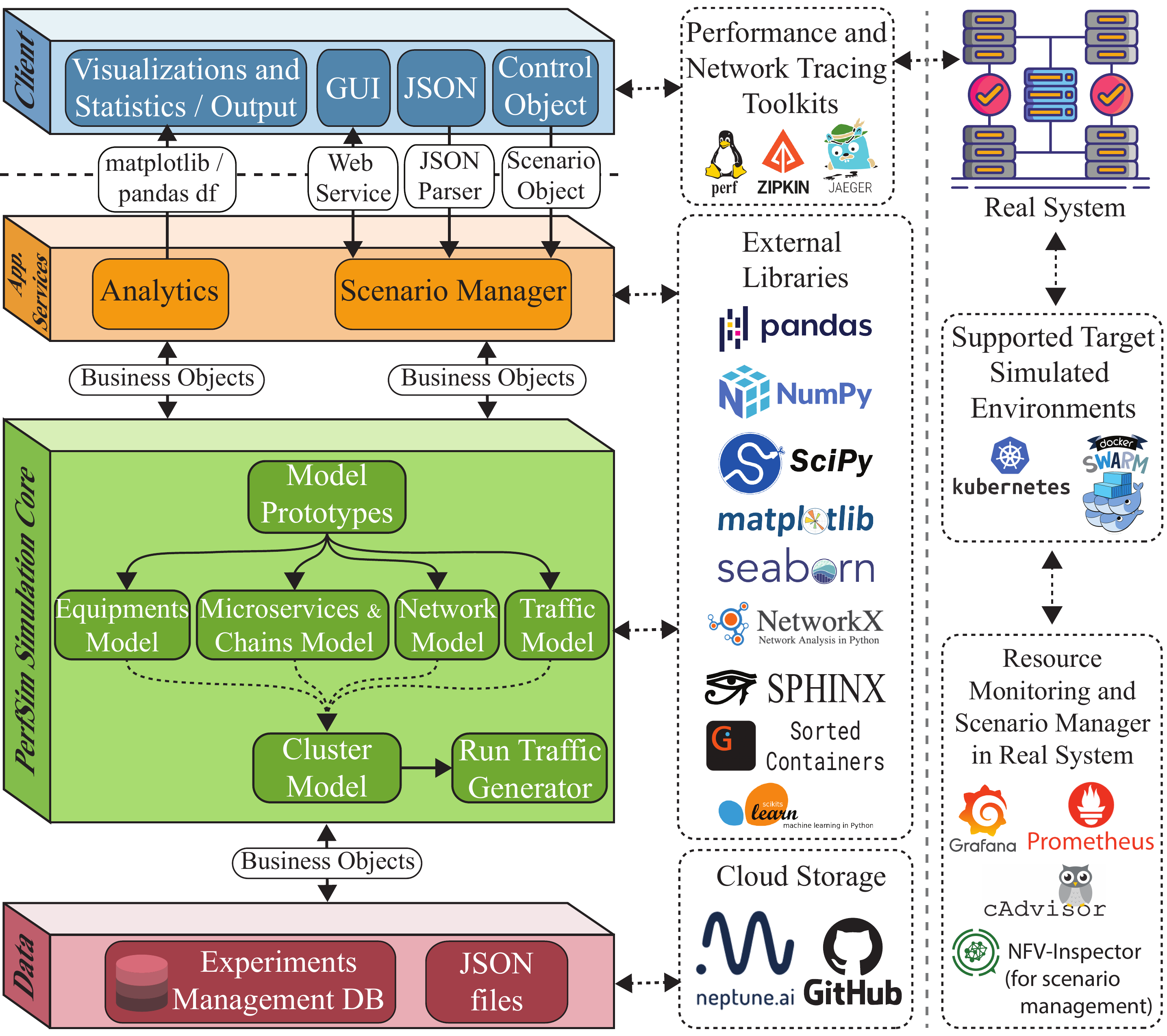}
	\caption{Software architecture of \name}
	\label{fig:software-architecture}
\end{figure}

Figure \ref{fig:software-architecture} presents the layered architecture of {\name}. We employed a multi-tier and object-oriented approach to architect {\name}: (1) The presentation layer (client), (2) business layer ({\name}'s core), and (3) data layer (results db).

\subsubsection{Presentation layer: defining scenarios}
To retrieve simulated performance insights using {\name}, a scenario should be specified for the simulator.

In the presentation layer, {\name} user defines the simulation scenarios and/or provides the control object. The user can provide the scenario in 3 general ways:

\begin{itemize}
    \item Using a JSON file for static simulations
    \item Using a \lstinline{perfsim.Cluster} object for dynamic simulations
    \item Using the GUI for quick experimentation
\end{itemize}

A scenario's structure in \name{} is as follows:

\begin{enumerate}
    \item \textbf{Prototypes} define the model of all microservices, endpoint function, threads, hosts, routers, links and traffics.
    
\begin{flushleft}
\begin{minipage}{\linewidth}
\begin{lstlisting}[
  caption=Prototypes JSON object in the scenario file,
  label={lst:prototypes},
  frame=tlrb,
  mathescape=true
]{Name}
"?\textbf{\color{blue}prototypes}?": {
 ?\indentrule?"microservices": {{$\smash{\forall{C_l}\!\in{\!C} \; \forall{\;^{l}\!S_i\!\in{\!^{l}\!S}} \; \forall{^{l}\!f_n^i\!\in{\!^{l}\!F_i}}\rightarrow}$ 
 ?\indentrule? ?\indentrule?"$\smash{\color{blue}^{l}\!f_n^i}$": [$\smash{\forall{t_m\in{^{l}\!f_n^i}}\rightarrow}$
 ?\indentrule? ?\indentrule? ?\indentrule? ?\indentrule? ?\indentrule? ?\indentrule? [$\smash{\color{darkorange}t_m^{\text{inst}},t_m^{\text{CPU}},t_m^{\text{maccs}},t_m^{\text{crefs}},t_m^{\text{cmiss}},t_m^{\text{cpenalty}},t_m^{\text{blk rw}}}$]]}},
 ?\indentrule?"hosts": {$\smash{\forall}$ host types $\rightarrow$
 ?\indentrule? ?\indentrule?"{type_name}": [$\smash{\color{darkorange}h^{\text{cores}},h^{\text{clock}},}${$\smash{\color{darkorange}\Hat{V}_H(r^H,*)\color{black}|r^H\in{R^H}}$}]},
 ?\indentrule?"routers": {$\smash{\forall}$ router types$\rightarrow$
 ?\indentrule? ?\indentrule?"{type_name}": [$\smash{\color{darkorange}\rho^{\text{lat}},\rho^{\text{in bw}},\rho^{\text{out bw}}}$]},
 ?\indentrule?"links": {$\smash{\forall}$ link types$\rightarrow$
 ?\indentrule? ?\indentrule?"{type_name}": [$\smash{\color{darkorange}l^{\text{lat}}}$]},
 ?\indentrule?"traffics": {$\smash{\forall}$ traffic types$\rightarrow$
 ?\indentrule? ?\indentrule?"{type_name}": [$\smash{\color{darkorange}C^{\text{rate}},C^{\text{duration}},C^{\text{batch}}}$]}}
      
\end{lstlisting}
\end{minipage}
\end{flushleft}

    \item \textbf{Equipments} define the hosts and routers that are going to be used in the cluster scenarios. These equipments should be created based on the defined "prototypes".

\begin{flushleft}
\begin{minipage}{\linewidth}
\begin{lstlisting}[
  caption=The JSON object defining hosts and routers,
  label={lst:equipments},
  frame=tlrb,
  mathescape=true,
]{Name}
"?\color{blue}\textbf{equipments}?": {
 ?\indentrule?"hosts": {$\smash{\forall{h_k\in{H}}\rightarrow}$ "$\color{blue}h_k$": "{host_type_name}"},
 ?\indentrule?"routers": {$\smash{\forall{\rho_{z}\in{P}}\rightarrow}$ "$\color{blue}\rho_z$": "{router_type_name}"}
\end{lstlisting}
\end{minipage}
\end{flushleft}

\item \textbf{Topologies} define the network topologies that are going to be used in the cluster scenarios.

\begin{flushleft}
\begin{minipage}{\linewidth}
\begin{lstlisting}[
  caption=Defining network topologies and hosts links,
  label={lst:topology},
  frame=tlrb,
  mathescape=true,
]{Name}
"?\color{blue}\textbf{topologies}?": {$\smash{\forall}$ topologies and ?links?$\rightarrow$
 ?\indentrule?"{topology_name}": {"nodes":{$\smash{P\cup{H}}$}, "edges":{$\smash{L}$}}}
\end{lstlisting}
\end{minipage}
\end{flushleft}

    \item \textbf{Service chains} define the chain of microsrevices as a graph (nodes and edges) including all link payloads.

\begin{flushleft}
\begin{minipage}{\linewidth}
\begin{lstlisting}[
  caption=Defining service chains graph \(G(C_l)\),
  label={lst:res_alloc_scenarios},
  frame=tlrb,
  mathescape=true,
]{Name}
"?\color{blue}\textbf{sfcs}?":{$\smash{\forall{C_l\in{C}}\rightarrow}$"$\smash{C_l}$": {"nodes":{$\smash{\!\!\!~^{l}\!F}$},"edges":{$\smash{\!\!\!~^{l}\!E}$}}}
\end{lstlisting}
\end{minipage}
\end{flushleft}
    
    \item \textbf{Resource allocation scenarios} consists of resource policy templates that later on can be assigned to each replica in a cluster scenario.

\begin{flushleft}
\begin{minipage}{\linewidth}
\begin{lstlisting}[
  caption=Defining resource allocation scenarios,
  label={lst:res_alloc_scenario},
  frame=tlrb,
  mathescape=true,
]{Name}
"?\color{blue}\textbf{res\_alloc\_scenarios}?": {$\smash{\forall}$ res. alloc. scenario$\smash{\rightarrow}$
 ?\indentrule?"{res_scenario_name}": {$\smash{\forall{r^S\in{R^S}}\!\!\rightarrow}$"$\smash{\color{blue}r^S}$": $\smash{\color{darkorange}\Hat{V}_S(r^S\!,*)}$}}
\end{lstlisting}
\end{minipage}
\end{flushleft}
    
    \item \textbf{Placement scenarios} define the algorithm, weights and configuration of the container scheduler. By default, {\name} uses the \textit{Least Allocated} placement algorithm.

\begin{flushleft}
\begin{minipage}{\linewidth}
\begin{lstlisting}[
  caption=Placement algorithms and related options,
  label={lst:placement},
  frame=tlrb,
  mathescape=true,
]{Name}
"?\color{blue}\textbf{placement\_scenarios}?": {$\smash{\forall}$ placement scenario$\rightarrow$ 
 ?\indentrule?"{placement_scenario_name}": 
 ?\indentrule? ?\indentrule?"algorithm": "{algorithm_name}", 
 ?\indentrule? ?\indentrule?"options": {e.g., $\smash{\forall{r^H\in{R^H}}}\rightarrow$ "$\smash{\color{blue}r^H}$": $\smash{\color{darkorange}W^{r^H}}$}}
\end{lstlisting}
\end{minipage}
\end{flushleft}

    \item \textbf{Affinity rule-sets} define the affinity/anti-affinity ruleset scenarios that can be used in the cluster scenario.

\begin{flushleft}
\begin{minipage}{\linewidth}
\begin{lstlisting}[
  caption=Rulesets containing affinity\(\!\)/anti-affinity rules,
  label={lst:affinity_rulesets},
  frame=tlrb,
  mathescape=true,
]{Name}
"?\color{blue}\textbf{affinity\_rulesets}?": {$\smash{\forall}$ ?affinity? ruleset$\rightarrow$
 ?\indentrule?"{affinity_set_name}": {
 ?\indentrule? ?\indentrule?"affinity": [$\smash{\forall{S_i\in{S}}\rightarrow{A_{S_i}}}$], 
 ?\indentrule? ?\indentrule?"?\color{blue}anti-affinity?": [$\smash{\forall{S_i\in{S}}\rightarrow{\widetilde{A}_{S_i}}}$]}}
\end{lstlisting}
\end{minipage}
\end{flushleft}
    
    \item \textbf{Cluster scenario} defines ultimate cluster scenarios and consists of a combination of all aforementioned sections, as well as additional configuration to define number of replicas for each service and network timeouts.

\begin{flushleft}
\begin{minipage}{\linewidth}
\begin{lstlisting}[
  caption=The ultimate simulation scenarios object,
  label={lst:ultimate_scenario},
  frame=tlrb,
  mathescape=true,
]{Name}
"?\color{blue}\textbf{cluster\_scenarios}?": {$\smash{\forall}$ cluster scenario$\rightarrow$
?\indentrule?"{scenario_name}": {
?\indentrule??\indentrule?"service_chains":{$\smash{\forall{C_l\in{C}}\rightarrow}$
?\indentrule??\indentrule??\indentrule?"$\color{blue}C_l$": { 
?\indentrule??\indentrule??\indentrule??\indentrule?"traffic_type": "{traffic_type_name}", 
?\indentrule??\indentrule??\indentrule??\indentrule??\indentrule?"nodes_settings": {$\smash{\forall{S_i\in{S}}\rightarrow}$
?\indentrule??\indentrule??\indentrule??\indentrule??\indentrule??\indentrule?"$\color{blue}S_i$": {
?\indentrule??\indentrule??\indentrule??\indentrule??\indentrule??\indentrule??\indentrule??\indentrule?"replica_count": $\color{darkorange}|S_i|$,
?\indentrule??\indentrule??\indentrule??\indentrule??\indentrule??\indentrule??\indentrule??\indentrule?"res_scenario": "{res_scenario_name}"}}}},
?\indentrule?"placement_scenario": "{placement_scenario_name}",
?\indentrule?"topology": "{topology_name}"}}
\end{lstlisting}
\end{minipage}
\end{flushleft}

\end{enumerate}

Similar to static JSON-based scenarios, a user can define the ScenarioManager object in pure Python and provide the control object to {\name} to start the simulation. As shown in Listing 1, since the JSON-based scenario is human-readable, a {\name} user can easily modify scenarios and/or add new ones as needed. Moreover, a user can define and simulate various types of scenarios without any need for performance trace logs or profiling data. The aforementioned scenarios can be provided based on either extracted performance models mentioned in the previous section or based on user-defined models. A {\name} user can then retrieve simulated insights of the final performance of the system in the given scenario.

\subsubsection{Business and data layers}
In the business layer, {\name} creates all necessary objects to perform the simulation. The details of modeling and simulation in {\name} is covered in the next section. 

After performing the simulation, the extracting performance simulation resutls will be saved in a database. The user can choose between saving the result in (1) a JSON file or (2) store it in a database (e.g., neptune.ai or MySQL).

{
\begin{algorithm}[htbp]
	\caption{Simplified placement of service replicas among several hosts in a cluster}
	\label{alg::placement}
	\SetKwProg{Fn}{function}{}{}
	\SetKwProg{ForEach}{foreach}{}{}
	\SetKwIF{If}{ElseIf}{Else}{if}{}{else if}{else}{}
	\SetKwFor{While}{while}{}{}%

	\small{

	\Fn{\texttt{filter\_hosts} ($\Hat{s}_j^i$)}{
	$A'_S\leftarrow{x \text{ in } A_{S_i}[x] \text{ if } A_{S_i}[x]\text{=}1}$

	$\widetilde{A}'_S\leftarrow{x \text{ in } \widetilde{A}_{S_i}[x] \text{ if } \widetilde{A}_{S_i}[x]\text{=}1}$

	$H^{\Hat{s}_j^i}\leftarrow{\text{ if } A'_S=\emptyset  \text{ then } [] \text{ else } H}$

	\ForEach{$S'\in{A'_S}$}{
		\ForEach{$s'\in{S'}$}{
			$H^{\Hat{s}_j^i}$.push($\Pi(s')$)
		}
	}

	\ForEach{$\xoverline{S}'\in{\widetilde{A}'_S}$}{
		\ForEach{$\xoverline{s}'\in{\xoverline{S}'}$}{
			$H^{\Hat{s}_j^i}$.pop($\Pi(\xoverline{s}')$)
		}
	}

	\ForEach{$h_k\in{H'}$}{
	\ForEach{$(r^S,r^H)\in{R: R^S\rightarrow{R^H}}$}{
	\If{$V_S(r^S,\Hat{s}_j^i)>V_H(r^H,h_k)$}
	{$H^{\Hat{s}_j^i}.\text{pop}(h_k)$}
	}

	}
	$\textbf{return } H^{\Hat{s}_j^i}$

	}

	\Fn{\texttt{score} ($\Hat{s}_j^i$,$h_k$)}{
		$\psi^{\text{min}}\leftarrow{0}, \psi^{\text{max}}\leftarrow{100},\psi_{\Hat{s}_j^i}^{h_k}\leftarrow{\text{based on Eq. \ref{eq:score}}}$

		$\textbf{return } \psi_{\Hat{s}_j^i}^{h_k}$
	}

	\Fn{\texttt{place} ($\Hat{s}_j^i$,$h_k$)}{
		\text{update\_values}($\Pi,V_H,V_S$)

	}

	\Fn{\texttt{schedule} ()}{
	\ForEach{$\Hat{s}_j^i\in{Q}$}{
	$H^{\Hat{s}_j^i}\leftarrow{\text{filter\_hosts}(\Hat{s}_j^i)}$
    
    \SetInd{0.5em}{0.9em}
	\If{$H^{\Hat{s}_j^i}=\emptyset$}{$Q.$pop$(\Hat{s}_j^i)$.push($\Hat{s}_j^i$)}
	$h^*\leftarrow{\text{first } h\in{H^{\Hat{s}_j^i}}},\xoverline{\psi}_{\Hat{s}_j^i}^{h_k}\leftarrow{\text{score(}\Hat{s}_j^i,h^*\text{)}}$
    
    \SetInd{0.5em}{0.5em}
	\ForEach{$h_k\in{H^{\Hat{s}_j^i}}$}{
	    \SetInd{0.5em}{0.9em}

		\If{$\xoverline{\psi}_{\Hat{s}_j^i}^{h_k}$\textgreater$\text{score}(\Hat{s}_j^i,h_k)$}{
			$h_{\Hat{s}_j^i}^*\leftarrow{\nth{1}\text{ } h\in{H^{\Hat{s}_j^i}}},\xoverline{\psi}_{\Hat{s}_j^i}^{h_k}\leftarrow{
			\text{ score(}\Hat{s}_j^i,h_{\Hat{s}_j^i}^*\text{)}}  $}
	}

	place($\Hat{s}_j^i,h_{\Hat{s}_j^i}^*$)

	$Q.$pop($S_j^i$)
	}
	}
	}
\end{algorithm}
}
\section{Placement, Chaining and Routing of Services over the Cluster}

In \name{}, the first step towards approximating the performance of service chains in a user-defined scenario is to place replicas on a cluster of hosts. After the placement, \name{} simulates the traffic scenario by first processing the defined service chain flow graphs, identifying parallel subchains, and extracting the exact route between service replicas based on the given network topology \(\tau\).

\subsection{Placement of service replicas on the cluster}

\begin{figure}[htbp]
    \centering \includegraphics[width=0.66\linewidth]{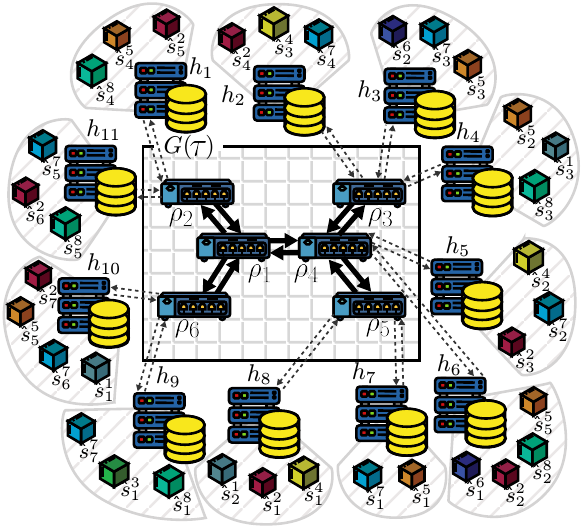}
    \captionof{figure}{An example placement of replicas over multiple hosts connected to a network with topology \(\tau\)}
  \label{fig:placement-and-topology}
\end{figure}

By default, we implemented a simplified version of the Least Allocated bin packing algorithm for the placement of service replicas, partially imitating the Kubernetes scheduler's placement strategy. The key idea behind this strategy is to ensure that replicas are placed on hosts with adequate resources and balance out the resource utilization of hosts. It consists of four steps: (1) enqueuing replicas, (2) filtering hosts, (3) scoring hosts, and (4) placement of the replica on the host. All these steps are managed by a \textit{schedule} procedure. Algorithm \ref{alg::placement} represents each of these steps and Figure \ref{fig:placement-and-topology} represents an example of such final placement.

\subsubsection{Enqueuing replicas}

Replicas are being placed one after the other. When a service requests for scheduling a replica on the cluster, scheduler adds that request to a queue (denoted as \(Q\) in Algorithm \ref{alg::placement}) and then attempts to find a suitable host for placing it.

\begin{equation}
	Q=\{\Hat{s}_j^i | \Hat{s}_j^i\in{S_i}\in{S}\}
\end{equation}

\subsubsection{Filtering hosts}

Considering all affinity/anti-affinity rules as well as re-source constraints of the service replica, the scheduler needs to filter out all the hosts that matches the specified limitations by the service. The filtered nodes, denoted as \(H^{\Hat{s}_j^i}\), are potentially eligible to host the replica \(\Hat{s}_j^i\in{S_i}\).

\begin{equation}
	H^{\Hat{s}_j^i}\text{=}\{h_k|\texttt{filter}\text{\_}\texttt{hosts}(\Hat{s}_j^i)\}
\end{equation}

\subsubsection{Scoring hosts}
In the Least Allocated bin packing strategy, in order to fairly distribute a replica \(\Hat{s}_j^i\) among several hosts in a cluster, all eligible hosts in the \(H^{\Hat{s}_j^i}\) are scored based on the request to capacity ratio of primary resources \(r^S\in{R^S}\) and \(r^H\in{R^H}\) considering the weight of each resource denoted as \(W^{r^H}\). The score \(\psi_{\Hat{s}_j^i}^{h_k}\) for each host \(h_k\) related to a replica \(\Hat{s}_j^i\) are between \(\psi^{\text{min}}\) and \(\psi^{\text{max}}\) are calculated as follows:

\begin{equation}
\label{eq:score}
	\psi_{\Hat{s}_j^i}^{h_k}=\frac{\psi^{\text{max}}\mathlarger{\mathlarger{\mathlarger{\sum}}}\limits_{\scalebox{0.5}{\((r^S,r^H)\in{R}\)}}{\bigg(1-\big(\frac{V_H(r^H,h_k)-V_S(r^S,
	\hat{s}_{j}^{i})}{{\Hat{V}_H(r^H,h_k)}}\Big)\bigg){W^{r^H}}}}{\mathlarger{\sum}\limits_{{r}^{H}\in{R^H}}{W^{r^H}}}
\end{equation}

\subsubsection{Placement of replicas}
After calculating the score of all eligible nodes for hosting replica \(\Hat{s}_j^i\), the host \(h_{\Hat{s}_j^i}^*\) with the lowest score will be selected:

\begin{equation}
	h_{\Hat{s}_j^i}^* = \Big\{h^*\in{H^{\Hat{s}_j^i}} \big| \psi_{\Hat{s}_j^i}^{h^*}=\text{max}\big\{\psi_{\Hat{s}_j^i}^{h_k} | \forall{h_k\in{H^{\Hat{s}_j^i}}}\big\}\Big\}
\end{equation}

\subsection{Service chain flow graphs}

A network flow graph \(G(C_l)=(\!\!~^{l}\!S,\!\!~^{l}\!E)\) of service chain \(C_l\) is a directed ordered graph that can have any form, from a

\begin{figure}[htbp]
    \centering \includegraphics[width=0.97\linewidth]{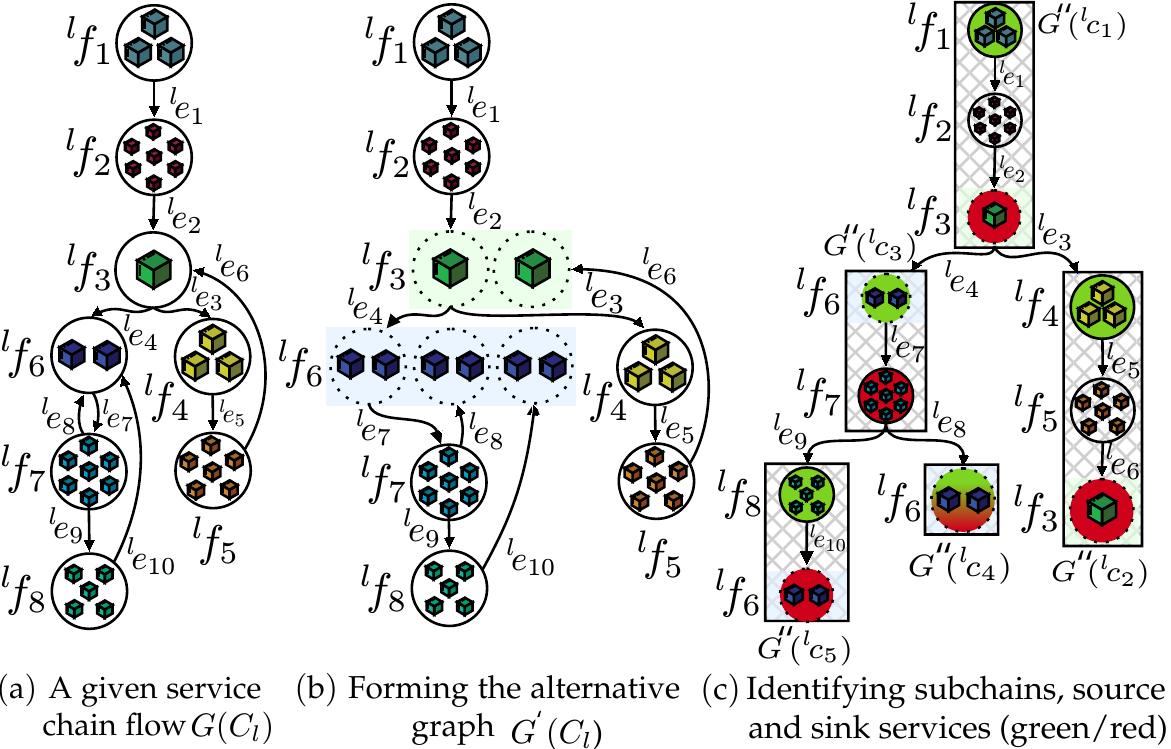}
    \caption{Forming alternative graph \(G'\) and flow graph \(G''\)}\label{fig:subchains}
\end{figure}

 \!\!\!\!\!\!\!\!\!simple tree-like service chain to a complex multi-degree cyclic multigraph. To efficiently simulate these service chains, we should be able to estimate the processing time of each service replica \(\!\!~^{l}\!\Hat{s}_j^i\in{\!\!~^{l}\!S_i}\in{\!\!~^{l}\!S}\subseteq{S}\), and then considering the outgoing request payload and network congestion, estimate the network transfer time between two services \((\!\!~^{l}\!S_i,\!\!\!\!~^{l}\!S_{i'})\in{\!\!~^{l}\!E}\). To accomplish these tasks, we should first identify the exact execution order of service requests.

In any given service chain \(C_l\), there might be subchains that run in parallel and their execution order are entirely depends on their execution time; this in turn depends on a vast range of parameters, starting from the processing power of the host to allocated resources to each service.

To identify all subchains within a service chain flow graph \(G(C_l)\), we first form an alternative graph \(G'(C_l)=(\!\!~^{l}\!S',\!\!~^{l}\!E',\!\!~^{l}\!F)\) from the original graph \(G(C_l)\) in which every service \(\!\!~^{l}\!S'_{i}\in{\!\!~^{l}\!S'}\) is visited only once. We form \(G'(C_l)\) by duplicating services in \(\!\!~^{l}\!S_i\in{\!\!~^{l}\!S}\) that are visited more than once; in other words, duplicate service nodes where has indegree of \(\delta^{-}(\!\!~^{l}\!S'_{i})\geq{2}\), and rerouting connecting edges to newly generated nodes. Then, based on newly formed \(G'(C_l)\), we now identify all \(n^{G'(C_l)}\) subchains in \(C_l\), denoted as \(\{\!\!\!~^{l}\!c_{x}\}_{x=1}^{n^{G'(C_l)}}\). Each subchain \(\!\!~^{l}\!c_{x}\) has a flow subgraph \(G^{''}(\!\!~^{l}\!c_{x})=(\!\!~^{l}\!S''_{x},\!\!~^{l}\!E''_{x},\!\!~^{l}\!F''_{x})\) that has a source service, which is the first service of the subgraph, and a sink service, which is the last service of the subgraph. In case a \(G^{''}(\!\!~^{l}\!c_{x})\) has only one service, that single service will be both source and sink node. We form each subgraph \(G^{''}(\!\!~^{l}\!c_{x})\) of the subchain \(\!\!~^{l}\!c_{x}\) by first identifying its source service. Starting from the the very first source service \(\!\!~^{l}\!S'_1\in{\!\!~^{l}\!S'}\) that is initiating first subchain \(\!\!~^{l}\!c_{1}\), when a service \(\!\!~^{l}\!S'_{i}\in{\!\!~^{l}\!S'}\) has an outdegree of \(\delta^{+}(\!\!~^{l}\!S'_{i})\geq{2}\), which means it has more than one outgoing edges \(\{e'^{l}_y=(\!\!~^{l}\!S'_{i}, \!\!~^{l}\!S'_x)\}_{\in{\!\!~^{l}\!E'}}\), then all immediate services \(\!\!~^{l}\!S'_x\) will be marked as source services and will initiate a new subchain. A subchain ends where we meet a service node with an outdegree of \(\delta^+(\!\!~^{l}\!S'_{y})\neq1\). A request \(\!\!~^{l}\!u_o\) ends when all sink service nodes conclude their executions.

As an example, Figure \ref{fig:subchains} demonstrates all aforementioned steps. In subfigure (a), we see an example flow graph \(G(C_l)\) of service chain \(C_l\) with all its services \(\!\!~^{l}\!S\) (circles) and their containing service replicas (cubes) together with all edges \(\!\!~^{l}\!E\) connecting services together (arrows) with the source service node \(S_1^l\). In this graph, we see node \(S_3^l\) initiate two parallel subchains and has a indegree \(\delta^{-}(S_3^l)=2\). Also, \(S_6^l\) forms a cycle with \(S_7^l\) with indegree \(\delta^{-}(S_6^l)=3\). In the second subgraph (b), we form the alternative graph \(G'(C_l)\) by duplicating \(S_3^l\) and \(S_6^l\) and rerouting connected edges \(e_{l,6}\), \(e_{l,8}\), and \(e_{l,10}\). In subfigure (c), we identify all \(n^{G'(C_l)}=5\) subchains \(\{\!\!~^{l}\!c_{x}\}_{x=1}^{5}\) and form \(\{G''(\!\!~^{l}\!c_{x})\}_{x=1}^{5}\).

\subsection{Routing requests}

When a subchain \(\!\!~^{l}\!c_{x}\) of a service chain \(C_l\) receives a user request \(\!\!~^{l}\!u_o\in{^l{U}}\), based on the \(G''(\!\!~^{l}\!c_{x})\), it'll be routed to one of the service replicas of the source service \(\!\!~^{l}\!S''^{x}_1\in{\!\!~^{l}\!S''^{x}}\subseteq{\!\!~^{l}\!S}\subseteq{S}\) that is chosen based on a round-robin load balancing algorithm (neglecting the session affinity possibility). If the chosen replica \(\Hat{s}_j^1\in{\!\!~^{l}\!S''^{x}_1}\) is either placed in the same host as the requester's service replica or if \(\!\!~^{l}\!S''^{x}_1\) is the source service replica \(S_1^l\in{\!\!~^{l}\!S}\) of the service chain \(C_l\), then, assuming \(\!\!~^{l}\!S''^{x}_1\) is not a sink service, after the request is being process in the \(s''^{1}_j\), it will simply get routed to the next replica inside the same host. But in case the next replica is not in the same host, then it needs to get routed to the destination replica based on the network topology graph \(G(\tau)\).

\input{tikz/routing}

Considering \(G(\tau)\) is a directed acyclic graph and there should be one-and-only-one active path carrying information from a source to destination in a network, we extract the path between all pairs of hosts and form a set of \(\binom{|H|}{2}\) subgraphs denoted as \(\{G(\tau)_{h_a,h_b}=(P_{h_a,h_b},L_{h_a,h_b})\}_{h_a,h_b\in{H}}\) where \(P_{h_a,h_b}\) is a set of routers between \(h_a\) and \(h_b\) that are connected using \(L_{h_a,h_b}\) links. As an example in Figure \ref{fig:placement-and-topology}, we have eleven hosts \(H=\{h_1,\dots,h_{11}\}\) that are connected together using six routers \(P=\{\rho_1,\dots,\rho_6\}\) with a tree topology \(\tau\) and network topology graph \(G(\tau)\). Figure \ref{fig:routing} represents the path from \(\Hat{s}_1^1\) to \(\Hat{s}_1^2\) based on the network topology \(\tau\) represented in Figure \ref{fig:placement-and-topology}.

\section{Approximating execution time of threads in a multi-core host}

The next step after identifying and deploying service chains and network routers is to estimate the execution time of each endpoint function \(\!\!~^{l}\!f_{n}^{i}\in{\!\!~^{l}\!F_l}\subseteq{\!\!~^{l}\!F}\) when a request arrives alongside other running processes in a host. In each cluster scenario, there exist a set of service replicas that each of them are capable of running a set of endpoint functions. When a request arrives, an endpoint function may propagate a few threads during its execution and each of its threads, when load-balances on a core's runqueue, may perform a few different tasks at a time. To be precise, a thread may perform one of the following tasks at any given time:

\begin{itemize}
	\item Execute a CPU intensive task on the processor
	\item Read/write bytes of data from/to memory/cache
	\item Read/Write a file from/to storage
	\item Send/receive packets of information over the network
	\item Be in the \textit{idle} mode
\end{itemize}

Additionally, other running threads inside a host have direct influence on the CPU time of an application because of the CPU scheduling mechanism in the operating system. For instance, Linux's CFS load-balance threads over dozens of CPU runqueues based on their load, CPU share, CPU quota and other parameters. Moreover, a typical multithreaded service have a main thread and a number of worker threads that coordinate with each other through synchronization primitives, that results additional overhead on execution time \cite{8401866}.

Even though predicting the exact execution time of a thread in a host can be very complex due to the aforementioned complexities, and considering that cycle-level simulation of multi-threaded services can be a very time-consuming procedure, we can predict a rough approximation of its run time by properly categorize type of running tasks in a host and approximate the execution time of each task by considering each thread's parent cgroups' resource constraints as well as other parallel active threads in the host. Consequently, this approximation may neglects a few performance factors such as  microarchitecture dependent variable (i.e.,CPU instruction sets and CPU architecture). However, during the modeling phase in a reference PTE, these effects will be measured and considered in the model, and therefore they contribute to the final approximation.

To approximate the total execution time of a thread, we categorize its tasks into 3 main types and accumulate measured values:

\begin{enumerate}
	\item Accumulated CPU instructions to be executed
	\item Accumulated stall cycles due to cache misses, memory access or block I/O
	\item Payload to be sent to the next service over the network
\end{enumerate}

\subsection{Load-balancing of Threads over CPU runqueues}
The first thing before being able to approximate the execution time of threads on a multi-core hardware, we need to estimate threads placement in a set of active CPU cores. This is crucial for the estimation, as the execution time is directly being affected by the other running threads in a core's runqueue. Therefore, we implemented a simplified version of Linux's CFS load-balancing algorithm \cite{10.1145/2901318.2901326} in \name{}. We presented the details of our implementation in Algorithm \ref{alg::cfs-load-balancing}. For simplicity, we assume there are only one NUMA node (one CPU socket with multiple cores/runqueues in it).

{
		\begin{algorithm}[tbp]
			\caption{Simplified CFS load-balancing algorithm in a host \(h_k\)}
			\label{alg::cfs-load-balancing}
            \DontPrintSemicolon
			\SetKwProg{Fn}{function}{}{}
			\SetKwProg{ForEach}{foreach}{}{}
			\SetKwIF{If}{ElseIf}{Else}{if}{}{else if}{else}{}
			\SetKwFor{While}{while}{}{}%
			\SetKwRepeat{Do}{do}{while}

			\small{
			sched\_domains = \((\text{NUMA},\text{pairs})\)

			sched\_groups[NUMA][0] = \(\{\text{core}\}_{\text{core}\in{h_k}}\)

			sched\_groups[pairs] = \(\{\{(\nth{1} \text{core in pair}, \nth{2} \text{core in pair}\}_{\text{core}\in{\text{core pairs}}}\}_{\text{core pairs}\in{h_k}}\)

			\Fn{\texttt{load\_balance\_threads} (\(h_k\))}{

			\ForEach{\(\text{current\_core}\in{\{1\dots{h_k^{\text{cores}}}\}}\)}{

			\ForEach{$\text{sd}\in{\{\text{sched\_domains}\}}$}{

				\(\text{first\_core}\leftarrow{ \texttt{first\_idle\_core}(sd)}\)

				\If{\(\text{first\_core} = \emptyset\)}{
					\(\text{first\_core}\leftarrow{\texttt{first\_core}(sd)}\)
				}
				\If{first\_core = current\_core}{
					\texttt{continue}}

				\ForEach{\(\text{sched\_group}\in{\text{sd}}\)}{
					\text{sched\_group}.\texttt{load}= avg(\(t_m^{\text{load}}\in{\text{sched\_group}}\))
				}

				\text{current\_sg}=
				          \texttt{current\_sched\_group}(current\_core)

				\text{bussiest\_sg}=\texttt{bussiest\_sched\_group}()

				\If{\text{bussiest\_sg}.\texttt{load} $ > $ {\text{current\_sg}.\texttt{load}}}{
					\Do{\text{load balance was not successful}}{
						\text{bussiest\_core}=
						\text{bussiest\_sg}.\texttt{next\_bussiest\_core}()\;
				\vspace*{-.35cm}
				\texttt{load\_balance}(\text{\footnotesize{current\_core}},\text{\footnotesize{bussiest\_core}})\;
				\vspace*{-.35cm}

					}
				}
			}
			}
			}
			}
		\end{algorithm}
	}

The CFS balances cores runqueues based on their \textit{load} which is a measure that is a combination of threads accumulated weights (CPU shares) and average CPU utilization. To estimated weight of a thread \(t_m\in{f_{n}^{i}}\) on a core, we first divide CPU shares of its parent service \(S_i\in{\!\!~^{l}\!S}\) with total number of its running threads, and then divide the result with sum of all CPU shares currently running on the core:

\begin{equation}
	t_m^\text{share}=\frac{S_i^{\text{CPU share}}}{|f_{n}^{i}|}
\end{equation}

Assuming \(\Hat{f}_{h_k}\) is the set of running threads placed on \(h_k\):

\begin{equation}
	\forall t_m\in{\Hat{f}_{h_k}} \rightarrow{}
	t_m^\text{weight}=\frac{t_m^\text{share}}{\sum\limits_{t_{j}\in{\Hat{f}_{h_k}}}{\!\!\!\!t_{j}^\text{share}}}
\end{equation}

To calculate load of a thread:

\begin{equation}
	t_m^\text{load}=\frac{t_m^\text{runnable sum}\times{t_m^\text{weight}}}{t_m^\text{runnable period}}
\end{equation}

In which, \(t_m^\text{runnable sum}\) is the amount of time that the thread was runnable and \(t_m^\text{runnable period}\) is the total time that the thread could potentially be running.

Additionally, the load-balancing procedure takes place by considering cache locality and its hierarchical levels called \textit{scheduling domains}. Within each scheduling domain, load-balancing occurs between a set of cores, called \textit{scheduling groups}. Since we assumed there is only one CPU socket in each host, there exist only two scheduling domains: NUMA node level and core-pair level. Algorithm \ref{alg::cfs-load-balancing} represents our simplified version of the CFS algorithm used in \name{}.

\subsection{Approximating CPU time}

Here we explain \name{}'s approach towards approximating CPU time of threads when co-located in a host.

\subsubsection{Calculating auxiliary CPU share of a thread}

When a thread co-locates with other threads on a multi-core machine, its execution time may be affected by the threads on the runqueue. Since we do not intend to simulate CPU time in a cycle-by-cycle basis, we approximate the effect of CPU bandwidth and CPU quota of threads by first defining an auxiliary CPU share for each thread, denoted as \(t_m^{\text{share ratio}}\). For simplicity, we assume threads run in either (1) the \textit{Best Effort} mode, which implies there is no CPU quota and CPU bandwidth defined, or (2) in the \textit{Guaranteed} mode where CPU shares are fixed and guaranteed. For every threads in a runqueue, we calculate \(t_m^{\text{share ratio}}\in(0,1024]\) as follows:

\begin{multline}
	\forall{t_m}\in{\Hat{f}_{h_k}}\text{ s.t. } t_m\in{f_n^i} \rightarrow{} \\
	t_m^{\text{share ratio}} = \begin{cases}
		\frac{t_m^{\text{share}}\times{1024}}{\sum\limits_{t_{j}\in{\Hat{f}_{h_k}}}{\!\!\!\!t_{j}^\text{share}}} & \text{ if } S_i^{\text{CPU quota}}\text{=}0                                          \\
		t_m^{\text{share}}                    & \parbox{2.5cm}{elif $t_m^{\text{share}} \leq 1024$} \\
		1024     & \parbox{2.5cm}{elif $t_m^{\text{share}} > 1024$}
	\end{cases}
\end{multline}

\subsubsection{Approximating cache miss rate based on cache stores and CPU size}

The cache miss rate of a single thread can get affected by various hardware dependent factors such as:
\begin{itemize}
    \item CPU L1, L2, and Last Level Cache (LLC) sizes
    \item Support of CPU Cache Allocation Technology (CAT) allowing software control over LLC allocation per process
    \item microarchiture-level details
\end{itemize}

The cache miss rate may also get affected by following significant software dependant factors:

\begin{itemize}
    \item Allocated CPU size of process
    \item Memory access rate of co-located threads in runqueue
\end{itemize}
 
 Since the focus of \name{} was to model the performance of microservices in merely software-oriented scenarios, such as container's placement, container/host affinity/anti-affinity, and resource allocation policies, we assumed a fairly similar microarchitecture and cache implementation details between the CPUs of PTE and the PE.

Hence, during the performance modeling phase, the effect of both (1) CPU size and  (2) memory access rate (store/loads) of co-located threads is measured and a logarithmic regression model is being trained to predict the excessive cache miss rate of each thread. We denote the first effect on \(t_m\in{f_{n}^{i}}\) as \(t_m^{\text{CMC}}\) and the second one as \(t_m^{\text{CMT}}\).

\begin{equation}
	t_m^{\text{CMC}}\text{=}a \ln(\frac{S_i^{\text{CPU limits}}}{|f_{n}^{i}|}) + b
\end{equation}

\begin{equation}
	t_{m}^{\text{CMT}}\text{=}a \ln\Bigg(\hspace{20pt}\nsum[1.2]_{\mathclap{\substack{t_j\in
	\text{runqueue of } t_m}}}t_j^{\text{maccs}}\hspace{8pt}\Bigg) + b
\end{equation}

We then apply these excessive penalties to the measured miss rate on the reference PTE as follows:

\begin{equation}
	t_m^{\text{miss rate}}=\frac{t_m^{\text{cmiss}}}{t_m^{\text{crefs}}}\times{(t_m^{\text{CMT}} + 1)}\times{(t_m^{\text{CMC}}+1)}
\end{equation}

\subsubsection{CPU time approximation}

We used the \textit{CPU Performance Equation} promoted in \cite{patterson1990computer} to approximate the CPU time of an isolated thread \(t_m\in{f_n^i}\) on an idle core (in the Best Effort mode):

\begin{multline}
	\text{Isolated CPU Time}(t_m)=\\t_m^{\text{inst}}({t_m^{\text{CPI}}} + \frac{t_m^{\text{maccs}}}{t_m^{\text{inst}}}\times{t_m^{\text{miss rate}}}\times{t_m^{\text{cpenalty}}} )\times\frac{1}{h_k^{\text{clock}}}
\end{multline}

Considering that \(t_m^{\text{inst}}\) is the accumulated number of instructions that thread \(t_m\) executes during its lifetime, it also includes additional instructions that thread performs due to mem accesses, cache misses, blkio r/w and net I/O.

\begin{equation}
	t_m^{\text{cycle penalty}}=\frac{t_m^{\text{maccs}}}{t_m^{\text{inst}}}\times{t_m^{\text{miss rate}}}\times{t_m^{\text{cpenalty}}}
\end{equation}

\begin{equation}
	^{\Hat{t}}t_m^{\text{relative share}} = \frac{t_m^{\text{CPI}}\times{^{\Hat{t}}t_m^{\text{share ratio}} }}{t_m^{\text{CPI}}+t_m^{\text{cycle penalty}}}
\end{equation}

Assuming \(^{\Hat{t}}t_{m}^{\text{inst}}\) is the remaining instructions at any given time \(\Hat{t}\), we approximated the CPU time of a thread on a CPU runqueue at any given time as follows:

\begin{equation}
	^{\Hat{t}}t_m^{\text{CPU time}} = \frac{^{\Hat{t}}t_m^{\text{inst}}\times{t_m^{\text{CPI}}}\times{\frac{1}{h_k^{\text{clock}}}}}{^{\Hat{t}}t_m^{\text{relative share}}}
\end{equation}

Therefore, given a time period \(\Delta{T}\geq{^{\Hat{t}}t_m^{\text{CPU time}}}\) (nanoseconds), we can estimate the executed instructions in that period as follows:

\begin{equation}
	^{\Hat{t}+\Delta{T}}t_m^{\text{inst}} = ^{\Hat{t}}\!\!t_m^{\text{inst}}-\frac{\Delta{T}\times{^{\Hat{t}}t_m^{\text{relative share}}}}{t_m^{\text{CPI}}\times{\frac{1}{h_k^{\text{clock}}}}}
\end{equation}

\subsection{Approximating the storage I/O time}

Assuming a thread \(t_m\) is placed on host \(h_k\) with accumulated reads/writes of \(t_m^{\text{blk rw}}\), we can approximate its storage I/O time as Equation \ref{eq:blkiotime}.

\begin{equation} \label{eq:blkiotime}
	t_m^{\text{blkio time}} = \frac{t_m^{\text{blk rw}}}{h_k^{\text{blkio bw}}}
\end{equation}

At any given time \(\Hat{t}\), the remaining blkio time is denoted by \(^{\Hat{t}}t_m^{\text{blkio time}}\).

\subsection{Approximating execution time of a thread}

As described in the previous section, the execution time of a thread consists of the CPU time, blkio rw time and idle time. Therefore, given a time \(\Hat{t}\), we can estimate the execution time of thread on a core's runqueue as follows:

\begin{equation}
	^{\Hat{t}}t_m^{\text{exe time}}=^{\Hat{t}}\!\!t_m^{\text{CPU time}}+^{\Hat{t}}\!\!t_m^{\text{blkio time}}+^{\Hat{t}}\!\!t_m^{\text{idle time}}
\end{equation}

\subsection{Approximating network transfer time}
When a service replica \(^l\!{\Hat{s}_x^i}\!\in{\!^{l}\!S_{i}\in{\!\!~^{l}\!S}}\) attempts to send \(\!\!~^{l}\!e_{v}^\text{payload}\) bytes of data to another service replica \(^l\!{\Hat{s}_y^j}\in{\!^{l}\!S_{j}\in{\!\!~^{l}\!S}}\), they are either placed on the same host and the network transfer time is equal to zero, or they placed in two different hosts \(h_a=\Pi(^l\!{\Hat{s}_x^i})\) and \(h_b=\Pi(^l\!{\Hat{s}_y^j})\). In the latter case, a request needs to traverse an ordered subgraph \(G(\tau)_{h_a,h_b}=(N_{h_a,h_b},L_{h_a,h_b})\)  where \(N_{h_a,h_b}\) consist of both hosts \(h_a\) and \(h_b\) and all routers in between them. Also, \(L_{h_a,h_b}\) is all the network links between two hosts, including host links and topology links. 

Depending on bandwidth usage of active requests flowing over a network link, the available bandwidth of a link changes over time. We therefore, denote the maximum available bandwidth of each link \(l_o\in{L_{h_a,h_b}}\) at any give time \(\Hat{t}\) as  $\theta(l_o)_{\Hat{t}}$ and calculate the maximum available bandwidth between all links in \(L_{h_a,h_b}\) as follows:
\begin{equation}
    \Theta(L_{h_a,h_b})_{\Hat{t}}=\text{min}(\{\theta(l_o)\}_{l_o\in{L_{h_a,h_b}}})
\end{equation}
. We then for any given time $\hat{t}$, calculate the request bandwidth between $^l\!{\Hat{s}_x^i}$ and $^l\!{\Hat{s}_y^j}$, denoted as $\theta(^l\!{\Hat{s}_x^i},^l\!{\Hat{s}_y^j})_{\hat{t}}$, as follows:

\begin{multline}
	\theta(^l\!{\Hat{s}_x^i},^l\!{\Hat{s}_y^j})_{\hat{t}}=\\	
	\text{min}\bigg(
	\Theta(L_{h_a,h_b})_{\hat{t}},V_S(\text{out bw},^l\!{\Hat{s}_x^i}),V_S(\text{in bw},^l\!{\Hat{s}_y^j})\bigg)
\end{multline}

To calculate the network time between 2 service replica \(^l\!{\Hat{s}_x^i}\) and \(^l\!{\Hat{s}_y^j}\), denoted as \(\omega(^l\!{\Hat{s}_x^i},^l\!{\Hat{s}_y^j})_{\hat{t}}\), we calculate the transfer time by dividing the payload size with available bandwidth, and then sum it up with routers and links latency as follows:

\begin{equation}
	\omega(^l\!{\Hat{s}_x^i},^l\!{\Hat{s}_y^j})_{\hat{t}}=\frac{^{\Hat{t}}e_{v}^\text{payload}}{\theta(^l\!{\Hat{s}_x^i},^l\!{\Hat{s}_y^j})_{\hat{t}}}+\!\!\!\sum_{\!\!\rho_{z}\in{P_{h_a,h_b}}}\!\rho_z^{\text{lat}}+\!\!\sum_{l_o\in{L_{h_a,h_b}}}\!\!l_o^{\text{lat}}
\end{equation}

\subsection{Putting all together: The simulation}

During each iteration, {\name} predicts the next event to simulate. We categorized events into five main categories:

\begin{itemize}
    \item \textbf{Request Generation} simulates the incoming traffic to each service chain by generating requests.
    
    \item \textbf{Threads Generation} checks for new queued threads and generates them.
    
    \item \textbf{Threads Execution Time Estimation} estimates the execution time of threads and checks whether their execution is going to end before the next transmission completes, or vice versa.
    
    \item \textbf{Threads Execution} estimates consumed instructions in all threads and repeat load-balancing among available cores in each host.
    
    \item \textbf{Network Transmissions}: Estimates the remaining payload in each active transmission in the network.
\end{itemize}

\section{Simulation Accuracy Evaluation}
\label{sav}

    In this section, we address the followings to highlight the accuracy, speed, applicability and significance of \name{}.

\begin{enumerate}
    \item \textbf{Potential threats to evaluation validity} (Section \ref{sec:potentialthreats}) to explain our approach for ensuring evaluation validity.
	\item \textbf{Performance modeling accuracy} (Section \ref{sec:accuracy}) to reflect and report the \name{'s} simulation error using a comprehensive set of prevalent scenarios (Table \ref{tab:scenarios}).
	\item \textbf{Execution time of \name{} prototype} (Section \ref{sec:time}) to report the amount of time required to simulate each scenario using an early-stage prototype of \name{}.
	\item \textbf{Simulating large scale service chains} (Section \ref{sec:largescale}) to demonstrate the applicability of \name{} for simulating large-scale service chains.
	\item \textbf{Challenges and limitations} (Section \ref{sec:discussion}) to highlight \name{'s} practical limitations as well as key challenges in simulating performance of computer systems.
\end{enumerate}

\subsection{Experimental setup}

\textbf{Cluster setup}. To evaluate the simulation error of service chains' execution time between \name{} and the real setup, we deployed a Kubernetes cluster using four physical machines (as compute nodes), and connected them through one or more Netgear 10Gbps routers (depending on the topology). 
All our servers are based on Intel Core i7 microarchitecture and their hardware details specified in Table \ref{table:model-specification}. The latency of routers for processing a packet is 7.3e5ns, and links latencies are 4.2e5ns.

\noindent \textbf{Workload}. To evaluate various scenarios described in this section, we used \textit{sfc-stress} \cite{sfc-stress} that is a customizable synthetic service chain benchmarking suit capable of generating different types of service chains with various types of CPU/memory/blkio/storage intensive as well as user-defined workloads. In sfc-stress, workloads are comparable to benchmark suites used for microservices, such as DeathStarBench \cite{10.1145/3297858.3304013}. They are similar to real-world cloud native microservices in various ways, such as capability of invoking any requested number of threads when their endpoint function being called, REST API based, written in Node.js, cloud native, deployable on both Kubernetes and Docker, fully parametric, and easily customizable. Using sfc-stress provided us with the flexibility of easily changing workloads type, size, threads count, execution duration, resource allocation settings per service (both vertically and horizontally), and most importantly design custom service chains and generate periodic automated requested based on given arrival rate (request per second). 

\noindent \textbf{Simulation parameters}. We used the pre-release prototype version (\texttt{alpha-0.1}) of \name{} for this evaluation. To drive the simulations, we first extracted performance traces of various endpoint functions in sfc-stress and fed it to \name{} as the simulation's initial parameters (Table \ref{table:model-specification}). We only set \(W^{\text{millicores}}\) and \(W^{\text{mem}}\) to \underline{1} and set other weights to \underline{0} (similar to Kubernetes settings).

\noindent \textbf{Placement of service replicas}. We used Kubernetes default kube-scheduler with weights specified in Table \ref{table:model-specification} to place and govern service replicas.

\subsection{Potential threats to evaluation validity}
\label{sec:potentialthreats}
As with any evaluation, potential threats exist to reduce its validity. We, therefore, identified several of these threats before starting our evaluation process and paid great attention to address them during our evaluation.

\subsubsection{Evaluation consistency}
During our initial experiments, we realized using only one experiment might not be representative of the evaluation scenarios performance behaviour. Thus, we repeated each experiment four times and calculated the average for each data point. In the end, we calculated moving average for both simulated and actual results as appropriate.

\subsubsection{Procedural rigor}
Collocating hosts in our testbed with other nodes in a cluster may introduce additional noise to the system. Moreover, we realized connecting various hosts to the same router may affect the performance of the network intensive scenarios. To mitigate this problem and prevent any possible noise affecting the experiments results, we isolated the entire testbed by locating hosts in a separate rack and making sure no other hosts are connected to the routers.

We also noted that keeping energy saving options in the BIOS may slightly affect the CPU performance by altering its clock frequency. To mitigate this effect, we disabled energy saving options in the BIOS and optimizied it for production.

\subsubsection{Comprehensiveness of evaluation scenarios}
Another threat to our evaluation was to neglect common scenarios used in today's systems. Nowadays a real distributed system may consist of different type of services, each with various resource demands, workload sizes, interconnections, and expected average traffic. Thus, to ensure the comprehensiveness of our evaluation scenarios when studying the accuracy of \name{}, we designed 6 category of scenarios, each focusing on a distinct aspect of \name{}. These categories, as presented in Table \ref{tab:scenarios}, includes CPU intensive, memory intensive, network intensive, scenarios with multiple replicas, and scenarios with multiple endpoint functions per service.

\subsection{Latency prediction accuracy}
\label{sec:accuracy}

In this section, we evaluated the accuracy of \name{} when simulating 104 different scenarios in 6 distinct category.

\subsubsection{Evaluating distinctive classes of scenarios}

\begin{figure*}[htbp]
	\centering
	\includegraphics[width=.8\linewidth]{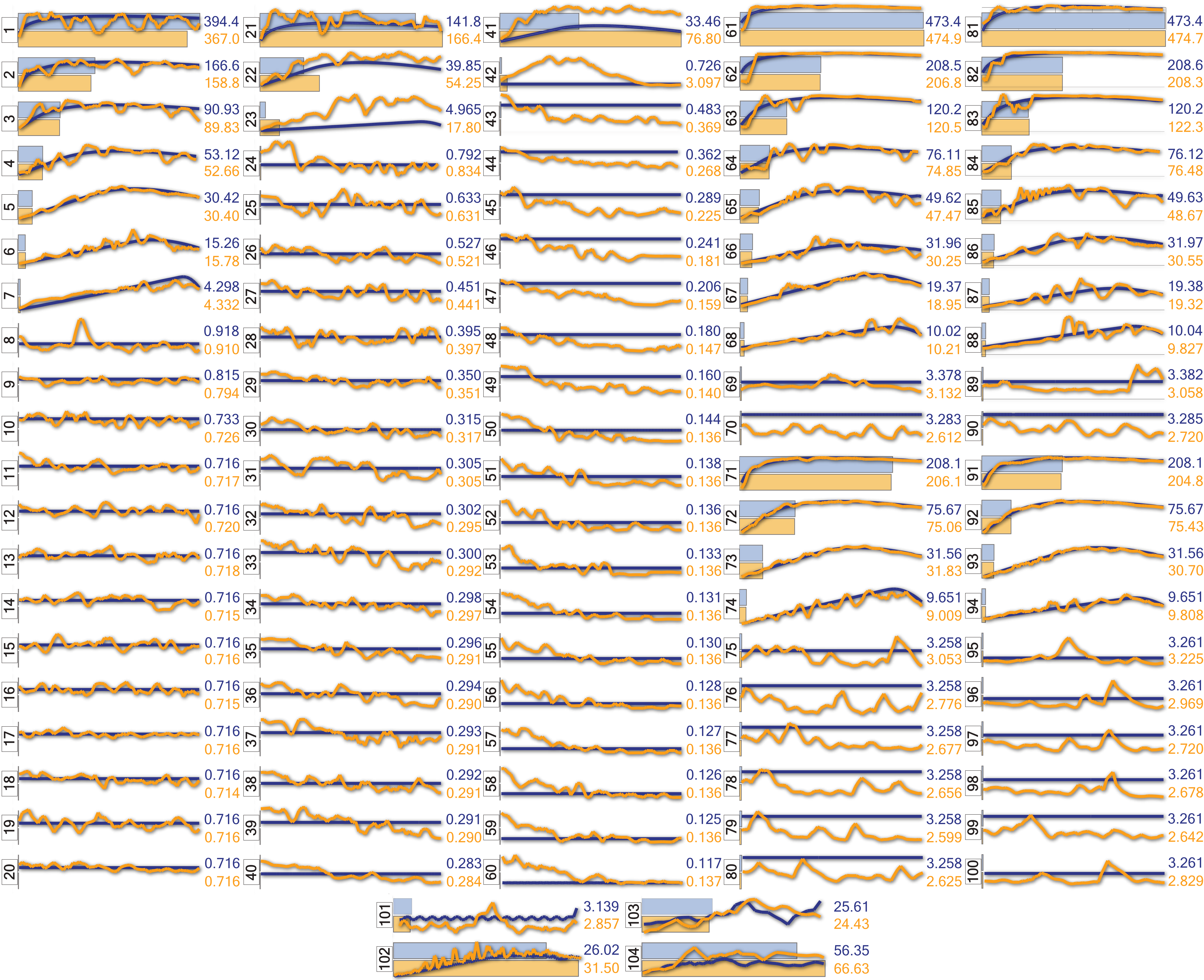}
	\caption{Evaluation results representing latency vs requests line graphs of \name{} (blue) and actual (orange) as well as the bar graph of their average latency (\(\overline{\!\!~^{l}\!u_o^{\text{exe time}}}\))}
	\label{fig:results}
\end{figure*}

We perform our evaluation by executing various workloads on our real Kubernetes cluster for \(\Delta\widehat{T}=60\) seconds and compared average execution time of requests (\(\overline{\!\!~^{l}\!u_o^{\text{exe time}}}\)) with the ones computed using \name{}. We summarized the simulation error of all scenarios in Table \ref{tab:scenarios} and illustrated the latency over requests line graphs in Figure \ref{fig:results}. We also illustrated the average latency (\(\overline{\!\!~^{l}\!u_o^{\text{exe time}}}\)) of each scenario as a bar graph in the same figure. We repeated each experiment four times to avoid any unaccounted artifact.

\begin{figure}[htbp]
 \includegraphics[width=.99\linewidth]{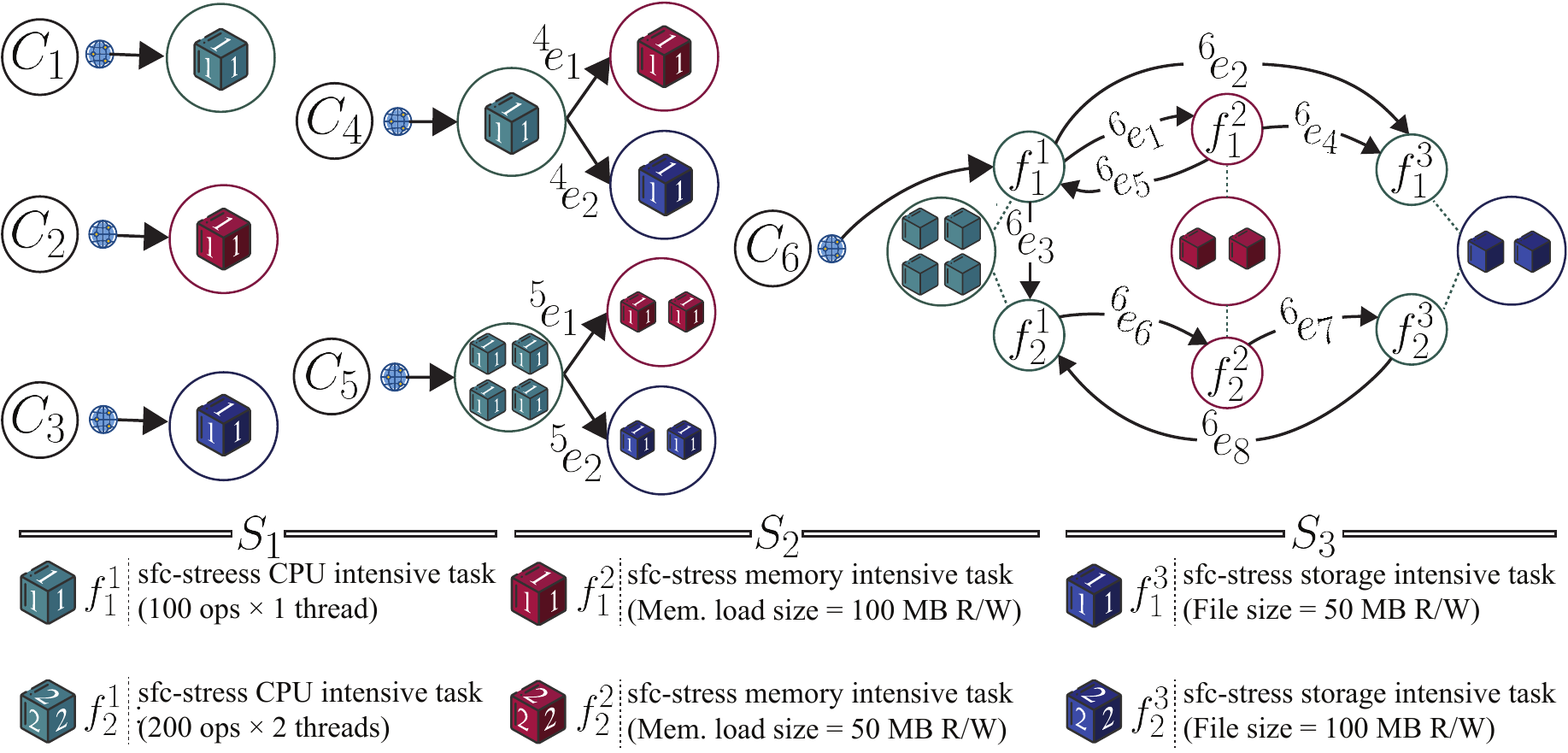}
            \caption{Service chain scenarios (\(\!\!~^{l}\!e_{v}^\text{payload}\)=50MB for all edges)}\label{fig:sfc}
    \end{figure}

\begin{figure}
     \centering
    \includegraphics[width=.9\linewidth]{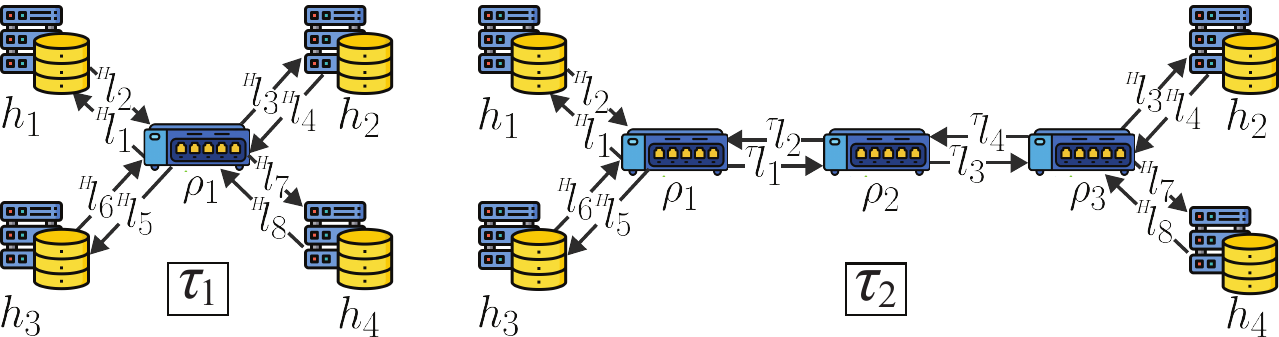}
           
            \caption{Network topologies used in the evaluation}\label{fig:topology}
    
        \end{figure}

Each scenario has (1) a service chain as illustrated in Figure \ref{fig:sfc}, (2) a network topology as illustrated in Figure \ref{fig:topology}, (3) an arrival rate \(C_l^{\text{rate}} \frac{\text{req}}{s}\), and (4) a resource allocation setting. 

In scenarios 1-20, our goal was to evaluate \name{} simulation accuracy when assigning different CPU sizes to a single CPU-intensive container. For example, in scenario\#1 we assigned 100 millicores to the only replica available in \(S_1\) and generated a traffic at a rate of 1 \(\frac{\text{req}}{\text{s}}\) on \(C_1\) and measured requests average execution time (\(\overline{\!\!~^{1}\!u_o^{\text{exe time}}}\)). We repeat the exact same scenario using \name{} and calculate the simulation percentage error by comparing extracted \(\overline{\!\!~^{1}\!u_o^{\text{exe time}}}\) in \name{} with the one measured in our real testbed.

\newdimen\parboxheight

\makeatletter
\newcommand*{\org@iiiparbox}{}
\let\org@iiiparbox\@iiiparbox
\renewcommand*{\@iiiparbox}[2]{%
	\ifx\relax#2%
		\setlength{\parboxheight}{0pt}%
	\else
		\setlength{\parboxheight}{#2}%
	\fi
	\org@iiiparbox{#1}{#2}%
}
\makeatother

\begin{table}[htbp]
	\setlength\tabcolsep{4.79pt}
	\setlength\doublerulesep{2pt}
	\setlength\extrarowheight{6pt}
	\caption{Constant control parameters of hosts, routers, links and workloads used in the evaluation}
	\label{table:model-specification}
	\begin{small}
	\begin{flushleft}
	\resizebox{1\columnwidth}{!}{%
		\begin{tabular}{|c|c|c|c|ccccccc|}
			\hhline{|*{11}{-}}
 			\hhline{|*{11}{-}}
            
			\hhline{|*{11}{-}|}
			\parbox[t]{3mm}{\multirow{4}{*}{\rotatebox[origin=c]{90}{\textbf{Hosts}}}} &
			\multirow{2}{*}{\(H\)} & \multicolumn{1}{c}{\multirow{2}{*}{\(h_k^{\text{clock}}\)}} & \multicolumn{1}{c}{\multirow{2}{*}{\(h_k^{\text{cores}}\)}} & \multicolumn{7}{c|}{\(h_k^{r^H}\)}\\
			\hhline{|~~~~*{7}{-}|}
			
			& & \multicolumn{1}{c}{} & \multicolumn{1}{c}{} & \(h_k^{\text{millicores}}\)&\multicolumn{2}{c}{\(h_k^{\text{mem}}\)}&\multicolumn{1}{c}{\(h_k^{\text{in bw}}\)}&\multicolumn{1}{c}{\(h_k^{\text{out bw}}\)}&\multicolumn{1}{c}{\(h_k^{\text{blkio bw}}\)} & \multicolumn{1}{c|}{\(h_k^{\text{blkio size}}\)} \\
            \hhline{|~|*{10}{-}|}
            \hhline{|~|*{10}{-}|}
			& \multirow{2}{*}{\parbox[M]{7pt}{\centering \(h_1\)\\\(\smash{\vdots}\)\\\(h_4\)}} &
			\multicolumn{1}{c}{\multirow{2}{*}{\parbox[M]{12pt}{\centering1.59\\\footnotesize GHz}}} & \multicolumn{1}{c}{\multirow{2}{*}{4}} & \multirow{2}{*}{4000} &
			\multicolumn{2}{c}{\multirow{2}{*}{\parbox[M]{13pt}{\centering16\\\footnotesize GB}}} &
			\multicolumn{1}{c}{\multirow{2}{*}{\parbox[M]{13pt}{\centering1\\\footnotesize Gbps}}} &
			\multicolumn{1}{c}{\multirow{2}{*}{\parbox[M]{13pt}{\centering1\\\footnotesize Gbps}}} &
			\multicolumn{1}{c}{\multirow{2}{*}{\parbox[M]{13pt}{\centering657\\\footnotesize MBps}}} &
			\multicolumn{1}{c|}{\multirow{2}{*}{\parbox[M]{13pt}{\centering500\\\footnotesize GB}}} \\
			\hhline{|*{11}{~}|}
			& & \multicolumn{1}{c}{} & \multicolumn{1}{c}{} & & & & & &\multicolumn{1}{c}{} & \multicolumn{1}{c|}{} \\
			\hhline{|*{11}{-}|}
			\hhline{|*{11}{-}|}
			\hhline{>{\arrayrulecolor{white}}*{11}{~}:>{\arrayrulecolor{black}}}
			\hhline{|*{11}{-}|}
			\hhline{|*{11}{-}|}
			\hhline{|-|-|*{9}{-}|}
             & \multirow{3}{*}{\(S\)} & \multicolumn{9}{c|}{Endpoint Functions}\\
			\hhline{|~|~|*{9}{-}|}
			& & \multirow{2}{*}{\(F\)} & \multicolumn{8}{c|}{Threads}
			\\
			\hhline{|~|~|~|*{8}{-}|}
			& & & \(T\) & \multicolumn{1}{c}{\(t_m^{\text{inst}}\)} & \multicolumn{1}{c}{\(t_m^{\text{cpi}}\)} & \multicolumn{1}{c}{\(t_m^{\text{maccs}}\)} & \multicolumn{1}{c}{\(t_m^{\text{crefs}}\)} & \multicolumn{1}{c}{\(t_m^{\text{cmiss}}\)} & \multicolumn{1}{c}{\(t_m^{\text{miss
			penalty}}\)} & \multicolumn{1}{c|}{\(t_m^{\text{blk rw}}\)} \\
			\hhline{|~|-|-|-|*{7}{-}|} 
			\hhline{|~|-|-|-|*{7}{-}|} 
			& & \(f_1^1\) &  \multicolumn{1}{|c|}{\cellcolor{white}\(t_{1,1,1}\)} & \multicolumn{1}{c}{$1.4e9$} & \multicolumn{1}{c}{0.7432} & \multicolumn{1}{c}{$3.1e8$} & \multicolumn{1}{c}{$1.0e6$} & \multicolumn{1}{c}{$1.0e5$} & \multicolumn{1}{c}{4} & \multicolumn{1}{c|}{\(\epsilon\)}\\
			\hhline{|~|~|-|>{\arrayrulecolor{gray!10}}~>{\arrayrulecolor{black}}|*{7}{~}|}
			 &  & \multirow{2}{*}{\(f_2^1\)} & \multicolumn{1}{|c|}{\cellcolor{gray!10} \(t_{1,2,1}\)} & \cellcolor{gray!10} $3.1e9$ & \multicolumn{1}{c}{\grey 0.750} & \multicolumn{1}{c}{\grey $7.2e8$} & \multicolumn{1}{c}{\grey $1.2e6$} & \multicolumn{1}{c}{\grey $1.3e5$} & \multicolumn{1}{c}{\grey 4} & \multicolumn{1}{c|}{\grey \(\epsilon\)}\\
			& \multirow{-3}{*}{\(S_1\)} & & \multicolumn{1}{|c|}{\cellcolor{white}\(t_{1,2,2}\)} & \multicolumn{1}{c}{$3.1e9$} & \multicolumn{1}{c}{0.715} & \multicolumn{1}{c}{$6.6e8$} & \multicolumn{1}{c}{$1.7e6$} & \multicolumn{1}{c}{$2.2e5$} & \multicolumn{1}{c}{3} & \multicolumn{1}{c|}{\(\epsilon\)}\\
			
			\hhline{|~|-|-|~|*{7}{~}|}
			& & \(f_1^2\) & \multicolumn{1}{|c|}{\cellcolor{gray!10}\(t_{2,1,1}\)} & \multicolumn{1}{c}{\cellcolor{gray!10}$1.7e9$} & \multicolumn{1}{c}{\cellcolor{gray!10}$0.520$} & \multicolumn{1}{c}{\cellcolor{gray!10}$3.4e8$} & \multicolumn{1}{c}{\cellcolor{gray!10}$2.9e6$} & \multicolumn{1}{c}{\cellcolor{gray!10}$2.0e6$} & \multicolumn{1}{c}{\cellcolor{gray!10}5} & \multicolumn{1}{c|}{\cellcolor{gray!10}\(\epsilon\)}\\
			\hhline{|~|~|-|>{\arrayrulecolor{gray!10}}~>{\arrayrulecolor{black}}|*{7}{~}|}
			& \multirow{-2}{*}{\(S_2\)} & \(f_2^2\) & \multicolumn{1}{|c|}{\cellcolor{white}\(t_{2,2,1}\)} & \multicolumn{1}{c}{$1.0e8$} & \multicolumn{1}{c}{0.4912} & \multicolumn{1}{c}{$7.4e8$} & \multicolumn{1}{c}{$5.5e6$} & \multicolumn{1}{c}{$4.1e6$} & \multicolumn{1}{c}{5} & \multicolumn{1}{c|}{\(\epsilon\)}\\
			\hhline{|~|-|-|~|*{7}{~}|}
			& & \(f_1^3\) & \multicolumn{1}{|c|}{\cellcolor{gray!10}\(t_{3,1,1}\)} & \multicolumn{1}{c}{\cellcolor{gray!10}$2.1e8$} & \multicolumn{1}{c}{\cellcolor{gray!10}0.6660} & \multicolumn{1}{c}{\cellcolor{gray!10}$4.3e7$} & \multicolumn{1}{c}{\cellcolor{gray!10}$1.5e6$} & \multicolumn{1}{c}{\cellcolor{gray!10}$5.7e5$} & \multicolumn{1}{c}{\cellcolor{gray!10}5} & \multicolumn{1}{c|}{\cellcolor{gray!10}$5.1e7$}\\
			\hhline{|~|~|-|>{\arrayrulecolor{gray!10}}~>{\arrayrulecolor{black}}|*{7}{~}|}
			\parbox[t]{2mm}{\multirow{-10}{*}{\rotatebox[origin=c]{90}{\textbf{Services}}}} & \multirow{-2}{*}{\(S_3\)}& \(f_2^3\) & \multicolumn{1}{|c|}{\cellcolor{white}\(t_{3,2,1}\)} & \multicolumn{1}{c}{$5.1e8$} & \multicolumn{1}{c}{0.7199} & \multicolumn{1}{c}{$2.2e7$} & \multicolumn{1}{c}{$4.3e6$} & \multicolumn{1}{c}{$2.3e6$} & \multicolumn{1}{c}{5} & \multicolumn{1}{c|}{$1.0e8$}\\
			
			\hhline{|*{11}{-}|} 
			\hhline{|*{11}{-}|} 
		\end{tabular}
		}
		\end{flushleft}
	\end{small}
\end{table}

In scenarios \#2-19, we increase the CPU size by 100 millicores in each scenario and repeat the same procedure; in scenario \#20, we also evaluated the case where all resources run in the best effort mode. In scenarios \#21-40, we repeated all aforementioned procedures  but for the memory intensive workloads (also implemented in the sfc-stress toolkit) to measure the accuracy of \name{} for simulating memory-intensive workloads. Similarly, in scenarios \#41-60, we focused on the disk-intensive workload.

To evaluate simulations involving networks and service chains, we designed scenarios \#61-80. In the first half of this category (scenarios \#61-70), we aimed to evaluate the accuracy of \name{} when assigning egress bandwidth to an outgoing service in service chain \(C_4\) (Figure \ref{fig:sfc}), and in the second half (scenarios \#71-80), our goal was to measure the accuracy when assigning ingress bandwidth to the incoming node \(S_3\). Scenarios \#81-100 were designed to measure the same effect but when the system is deployed in slightly more complex network topology \(\tau_2\) (Figure \ref{fig:topology}).

	\capbtabbox{%
		\setlength\extrarowheight{3pt}
        \raggedright
    	\setlength\tabcolsep{2pt}

    	\resizebox{.985\columnwidth}{!}{%

    		\begin{tabular}{cccccccc}
    			\toprule
    			\multirow{2}{*}{\parbox[t]{13mm}{\centering{}Scenario\\Type}}& \multirow{2}{*}{\parbox[t]{13mm}{\scriptsize\centering{}Experiment\\\normalsize{}Number}} &  \multirow{2}{*}{\(C_l^{\text{rate}}\) (\(\frac{\text{req}}{\text{s}}\))} & \multirow{2}{*}{\parbox[t]{26mm}{\centering{}Target Resource\\Capacity}} & \multirow{2}{*}{\(\tau\)} & \multicolumn{2}{c}{\(\overline{\!\!~^{l}\!u_o^{\text{exe time}}}\)} & \multirow{2}{*}{\parbox[t]{6mm}{\centering{}Avg.\\Error}}\\
    			\cmidrule{6-7}
    			& & & & & \footnotesize\textbf{{\name}} & \footnotesize\textbf{Actual}\normalsize & \\
    			\midrule
    			\multirow{2}{*}{\parbox[t]{13mm}{\centering{}CPU\\intensive}}& 1-19 & \multirow{-1}{*}{\(C_1^{\text{rate}}\)\shorteq{}1} & \(S_1^{\text{\footnotesize{}CPU requests}}\)\shorteq{}\(\{100k\}_{k\shorteq{}1}^{19}\) & \(\tau_1\) &  & & \\
    			& 20 & \multirow{-1}{*}{\(C_1^{\text{rate}}\)\shorteq{}1} & Best Effort & \(\tau_1\) & \multirow{-1}{*}{38.23} & \multirow{-1}{*}{36.42} &\multirow{-1}{*}{1.237\%}\\
    			\rowcolor{gray!10}
    			 & 21-39 & \multirow{-1}{*}{\(C_2^{\text{rate}}\)\shorteq{}1} & \(S_1^{\text{\footnotesize{}CPU requests}}\)\shorteq{}\(\{100k\}_{k\shorteq{}1}^{19}\) & \(\tau_1\) & & &\\
    			\rowcolor{gray!10}
    			\multirow{-2}{*}{\parbox[t]{13mm}{\centering{}Memory\\intensive}} & 40 & \multirow{-1}{*}{\(C_2^{\text{rate}}\)\shorteq{}1} & Best Effort & \(\tau_1\) & \multirow{-1}{*}{9.654} & \multirow{-1}{*}{12.24} &\multirow{-1}{*}{6.662\%}\\
    			\multirow{2}{*}{\parbox[t]{13mm}{\centering{}Storage\\intensive}} & 41-59 & \multirow{-1}{*}{\(C_3^{\text{rate}}\)\shorteq{}1} & \(S_3^{\text{\footnotesize{}CPU requests}}\)\shorteq{}\(\{100k\}_{k\shorteq{}1}^{19}\) & \(\tau_1\) & & &\\
    			& 60 & \multirow{-1}{*}{\(C_3^{\text{rate}}\)\shorteq{}1} & Best Effort & \(\tau_1\) & \multirow{-1}{*}{1.877} & \multirow{-1}{*}{4.144} &\multirow{-1}{*}{19.27\%}\\
    			\rowcolor{gray!10}
    			& 61-70 & \multirow{-1}{*}{\(C_4^{\text{rate}}\)\shorteq{}1} & \(S_1^{\text{\footnotesize{}out bw}}\shorteq{}\{100k\}_{k\shorteq{}1}^{10}\) & \(\tau_1\) & 99.60 & 98.97 &5.086\%\\
    			\rowcolor{gray!10}
    			& 71-80 & \multirow{-1}{*}{\(C_4^{\text{rate}}\)\shorteq{}1} & \(S_3^{\text{\footnotesize{}in bw}}\)\shorteq{}\(\{100k\}_{k\shorteq{}1}^{10}\)& \(\tau_1\) & 34.45 & 33.84 &12.77\%\\
    			\rowcolor{gray!10}
    			& 81-90 & \multirow{-1}{*}{\(C_4^{\text{rate}}\)\shorteq{}1} & \(S_1^{\text{\footnotesize{}out bw}}\)\shorteq{}\(\{100k\}_{k\shorteq{}1}^{10}\) & \(\tau_2\) & 99.62 & 99.60 &4.306\%\\
    			\rowcolor{gray!10}
    			\multirow{-4}{*}{\parbox[t]{13mm}{\centering{}Network\\intensive}} & 91-100 & \multirow{-1}{*}{\(C_4^{\text{rate}}\)\shorteq{}1} & \(S_3^{\text{\footnotesize{}in bw}}\)\shorteq{}\(\{100k\}_{k\shorteq{}1}^{10}\) & \(\tau_2\) & 34.45 & 33.78 &9.772\%\\
    			\footnotesize Multi-replica & 101-102 & \multirow{-1}{*}{\(C_5^{\text{rate}}\)\shorteq{}\(\{1,3\}\)} & Best Effort & \(\tau_1\) & 14.57 & 17.17 &13.63\%\\
    			\rowcolor{gray!10}
    			\scriptsize Multi-endpoint & 103-104 & \multirow{-1}{*}{\(C_6^{\text{rate}}\)\shorteq{}\(\{\frac{1}{2},\frac{1}{3}\}\)} & Best Effort & \(\tau_1\) & 40.98 & 45.53 &10.10\%\\
    			\bottomrule
    			& & & & & \multicolumn{3}{r}{Avg. error=9.203\%}
    		\end{tabular}
		}
	}{%
	\caption{Evaluation scenarios and associated errors}%
	\label{tab:scenarios}
	}

To evaluate scenarios involving multiple replicas, we designed scenarios \#101-102. In these scenarios, we assigned 4 replicas to \(S_1\), 2 replicas to \(S_2\), and 2 replicas to \(S_3\). We tested 2 different arrival rates on \(C_5\): (a) 1 \(\frac{\text{req}}{\text{s}}\) in scenario \#101, and (b) 3 \(\frac{\text{req}}{\text{s}}\) in scenario \#102. In scenarios 103-104, we aimed to evaluate deployment that involved multiple endpoint functions, as well as, cases where an endpoint function spawns multiple threads (e.g., \(f_2^1\)). Another goal of these scenarios was to test the performance of \name{} in scenarios with complex service chains (i.e., \(C_6\)) with multiple loops and nested diamond-shaped connections.

\subsubsection{Reflections on simulation accuracy}

In the vast majority of our experiments, as shown in Figure \ref{fig:results}, \name{} predicted the latency trend with a high accuracy (with average error rate of ~9\% error). However, in all scenarios, the actual service latency has visible fluctuations across time (the orange line) while \name{}'s predictions (blue line) did not capture those slight variations. This is because \name{} is a discrete-event simulator and captures changes in the system state based on events such as request generation/conclusion, thread spawn/kill, network transmission start/end, queue start/end, task scheduling start/end, etc. Therefore, as \name{} events are not defined based on the CPU clock resolution (i.e., kernel clock) to maintain fast simulation speed, it doesn't consider the CPU noise-factor to simulate slight fluctuations (e.g., slight changes in CPU frequency or clock speed).

Another observation captured from the experiments results is the clear harmony between almost all simulation results and the reality. Nonetheless, in some intense scenarios (e.g., scenarios 21-23 or 41-42)  where very low CPU resources has been assigned to a memory intensive or storage intensive services (100-300 millicores), even though \name{} accurately simulated the trend of performance, we observe a gap between \name{}'s  approximated latency and the actual obtained latency in reality. The nature of this variation is due to the use of a static model for each service in our experiments. In highly overloaded or intense scenarios, specially in a memory-intensive task, cache-miss rate may significantly affected and undergo substantial changes, which even though has already been considered in \name{}'s core, combining it with ever increasing rate of context switches and CPU migrations in such intense scenarios, creates a high competition between service threads from one hand and system/user threads on the other hand which puts the entire system in a chaotic situation. This problem can be addressed by utilizing more dynamic performance models per service and considering system level threads in the simulation.

\subsection{Execution time of \name{}}
\label{sec:time}

One of the key ideas behind designing \name{} is to eliminate the barrier for utilizing advanced machine learning techniques (such as deep reinforcement learning) for optimizing the performance of large-scale service chains that require fast efficiency in estimating various resource allocation and placement scenarios. In this section, we evaluated the execution time of all 104 scenarios and compared them with the simulation time in \name{}.

Due to the fact that the main users of \name{} will be researchers and performance engineers with limited access to computation resources, and to highlight the lightweight nature of \name{}, a single personal laptop has been used to measure the simulation speed. We ran \name{} in single-threaded mode using a MacBook Pro with a 2.6 GHz Intel Core i7 CPU and 16GB of RAM.

\begin{figure*}[htbp]
	\centering
	\includegraphics[width=.9\linewidth]{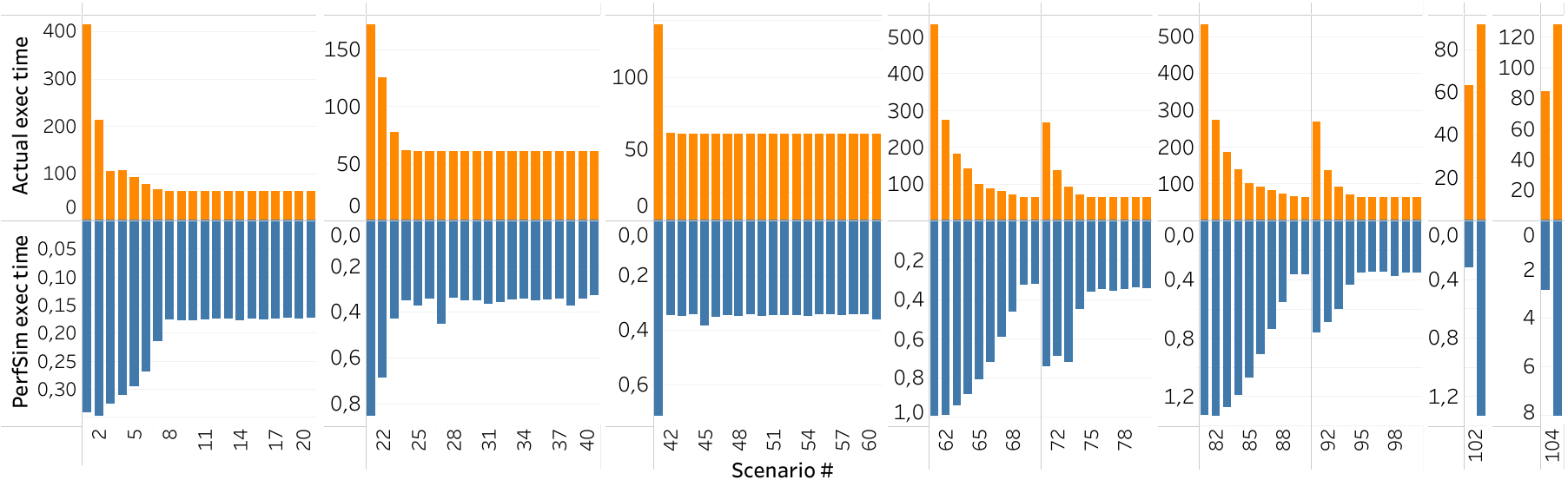}
	\caption{Execution time (in seconds) of all scenarios in both \name{} and real deployment (note the difference in scales).}
	\label{fig:execution-time}
\end{figure*}

We represents the detailed speed comparison in Figure \ref{fig:execution-time}. As shown in the graph, a single-thread execution of \name{} is 16-1200 times faster to evaluate the performance of a scenario when compared to running/evaluating the same deployment on a real cluster. 

\subsection{Simulating large scale service chains}
\label{sec:largescale}

\thisfloatsetup{heightadjust=all,valign=t}
\begin{figure}[htb]
\captionsetup[subfigure]{justification=centering}

    \ffigbox[]{
    \begin{subfloatrow}
        \ffigbox[.4\textwidth]
            {\centering\includegraphics[width=.92\linewidth]{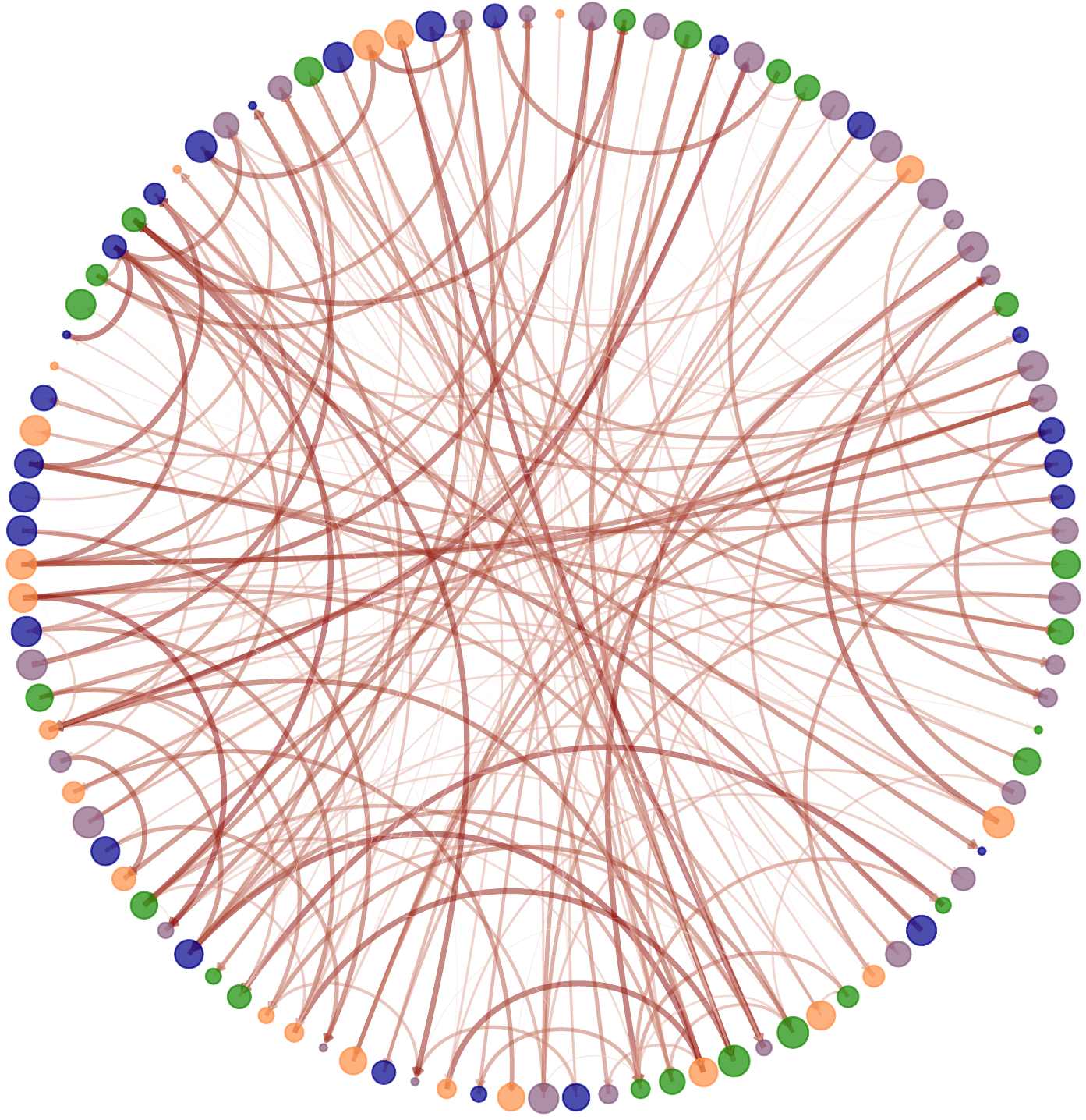}}
            {\subcaption{Original service chain \(G(C_l)\)}\label{fig:big-sfc-original-graph}}
    \end{subfloatrow}
    \begin{subfloatrow}
    \ffigbox[.59\textwidth]
        {\begin{flushright}\includegraphics[width=.89\linewidth]{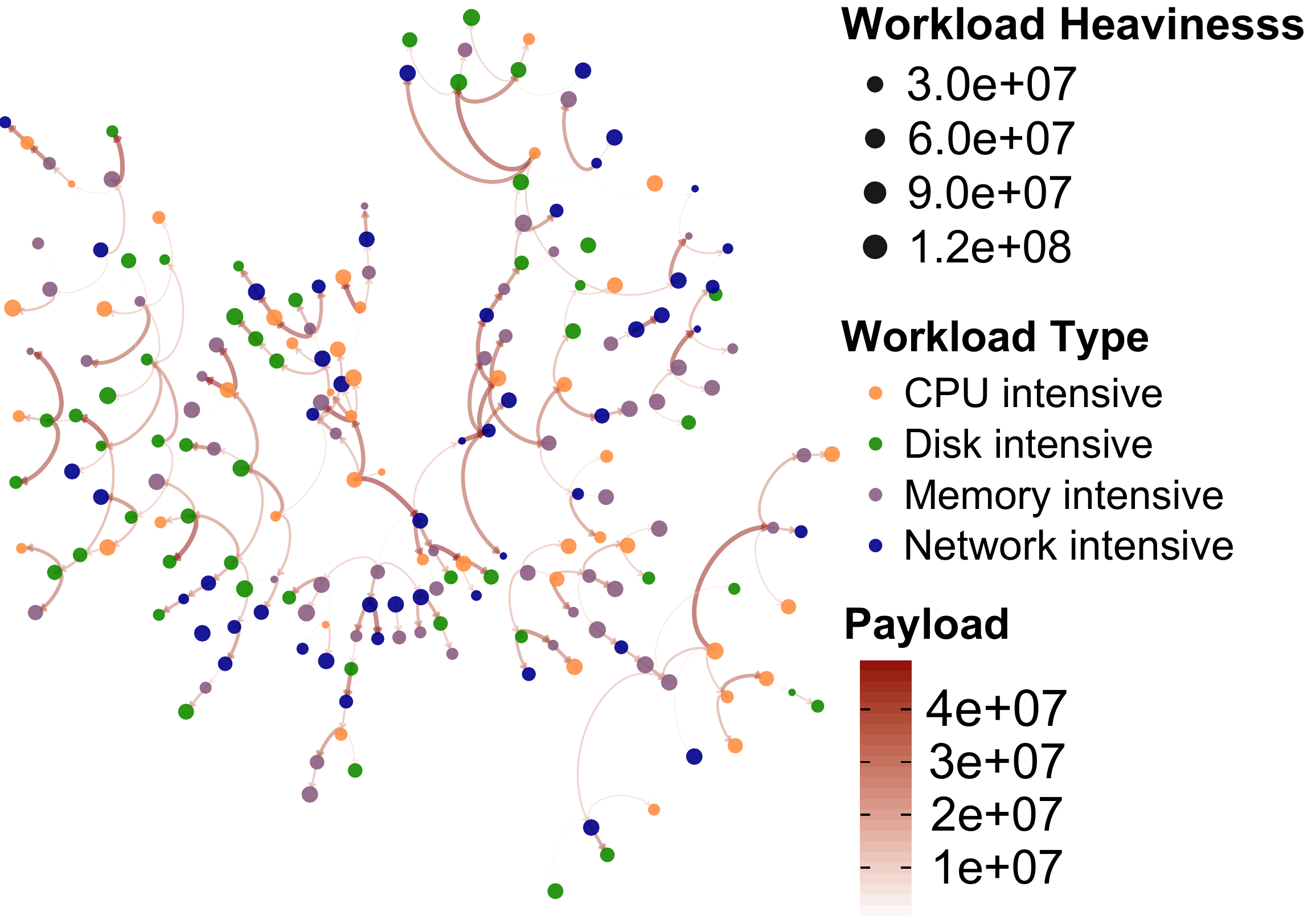}\end{flushright}}
        {\subcaption{Extracted alternative graph \(G'(C_l)\)}\label{fig:big-sfc-alt-graph}}
    \end{subfloatrow}
    }
  {\caption{Simulating a large service chain with different types of workload, payload sizes and heaviness (e.g., large number of instruction, cache r/w)}
  \label{fig:big-sfc}}
\end{figure}

When designing \name, we initially aimed to use it for optimizing large-scale service chains. Therefore, one of the aspects of this evaluation has been focused on the ability of \name{} to simulate large-scaled service chains. For this purpose, we simulated the performance of a large randomized service chain deployed over a cluster of 100 host. We illustrated both our generated service chain \(G(C_l)\) and extracted alternative graph \(G'(C_l)\) in Figure \ref{fig:big-sfc}. This service chain consists of 100 nodes and 200 edges with randomized payloads, workload heaviness and workload type.

Similar to the previous scenarios, we used a single-threaded version of \name{} for the entire experiment. We presents both simulation parameters and results in Table \ref{tab:big-sfc-results}. As shown, \name{} was able to simulate the performance of this large-scale service chain for up to 10 times faster than targeted execution time \(\Delta{\Hat{T}}\) while using reasonable CPU and memory resources.

\capbtabbox{%
        \raggedright
    	\setlength\tabcolsep{7.5pt}

	\resizebox{0.95\columnwidth}{!}{%

    		\begin{tabular}{cc||cc}
    			\toprule
    			\rowcolor{gray!20}
    			\multicolumn{2}{c}{\normalsize \textbf{Simulation parameters}} &   \multicolumn{2}{c}{\normalsize \textbf{Simulation results}}\\
    			\cmidrule{1-2} \cmidrule{3-4}
    			Parameter & Value & Key & Value  \\
    			\midrule
    			\rowcolor{gray!10}
    			\(C_l^{\text{rate}}\) & 0.1 \(\frac{\text{req}}{\text{s}}\) &\(\overline{\!\!~^{l}\!u_{o}^{\text{exe time}}}\) & 9,744.95ms\\
    			\(\Delta{\Hat{T}}\) & 600s & Sim. time & 68.80s \\
    			\rowcolor{gray!10}
    			\(|\!\!~^{l}\!S|\) & 100 & \(|\!\!~^{l}\!S'|\) & 201 \\
    			\(|\!\!~^{l}\!E|\) & 200 &  \(n^{G'(C_l)}\) & 161 \\
    			\rowcolor{gray!10}
    			\(|H|\) & 100 & Exec. threads & 12060 \\
    			\(\tau\) & \(\tau_1\) & CPU usage & 99.9\% \\
    			\rowcolor{gray!10}
    			Resources & Best effort & Mem. usage & 1.64GB
    		\end{tabular}
		}
	}{%
	\caption{Simulation parameters and summary of results in the large service chain scenario using \name{} prototype}%
	\label{tab:big-sfc-results}
	}

\subsection{Discussion on challenges and limitations}
\label{sec:discussion}

In this section, we discuss main challenges of performance simulation, as well as, key limitations for using \name.

\subsubsection{Comments on time-predictability of tasks}
Time-predictability of a task in a computer software mainly refers to the properties of the phenomenon execution time, including the execution pattern of instructions or spectrum of occurrence events related to a job's execution time \cite{10.1007/978-3-642-16256-5_5,article-time-predictability}. These properties directly affect the process of modeling a job’s execution time and, in many cases, can make it impossible to extract any reusable pattern \cite{6404196}. Hence, it is a vital property to hold in real-time and embedded systems due to their time and mission critical nature.

To shed light on the surface of the problem, we provide an example in Listing 2, representing 2 functions (\texttt{f1} and \texttt{f2}), both calculating \texttt{n} number of MD5 hashes. Function \texttt{f1} receives the value of \texttt{n} as an argument whereas \texttt{f2} initiates \texttt{n} with a randomly generated number between 0 and \(10^{10}\). For these functions, modeling \texttt{f1} is relatively straightforward because its execution time is directly related to the input argument \texttt{n}; for \texttt{f2} on the other hand, it is extremely hard to approximate the execution time because it depends on the randomly generated variable \texttt{n}.

\begin{flushleft}
\begin{minipage}[t]{0.49\linewidth}
\begin{lstlisting}[language=JavaScript]
//high time-predictability
function f1(n){
  for($i=0; $i<$n; $i++){
    createHash('md5');
  }
}

\end{lstlisting}
\end{minipage}%
\hfill\hfill\vrule\hfill
\begin{minipage}[t]{0.51\linewidth}
\raggedright
\begin{lstlisting}[language=JavaScript]
//low time-predictability
function f2(){
  let n = randInt(0,10**10);
  for($i=0; $i<$n; $i++){
    createHash('md5');
  }
}
\end{lstlisting}
\end{minipage}%
\captionof{lstlisting}[Foo]{An example of time-predictability in functions}
\vspace{\baselineskip}
\end{flushleft}

With the growing convergence of High-Performance Computing (HPC) systems and cloud computing paradigm, modern design patterns tend to progressively consider the properties of time-predictable computing in their approaches to be able to plan for the underlying resource allocation policies, infrastructure costs, as well as, to avoid Service Level Agreement (SLA) penalties. However, many organizations are still not ready for such a transition, and this imposes immense challenges in modeling and simulating the performance of their designed systems.

\subsubsection{Modeling the software architecture}
The software architecture, as the key foundation of any software system, can significantly impact the performance of chain of services. As described in the previous sections, \name{} can adequately model architecture of a software system by defining the connection between microservices and hosts, and translating each microservice into a set of endpoint functions while describing their properties.

However, one of the critical challenges in simulating large-scale systems is to model components that may change their behaviour based on the context. For example, some architectures foresee an HTTP cache server in the front-line of their main webserver to reduce the latency of requests or API calls. When a request arrives, these cache servers check their temporary storage for the requested content and only send the request to the upstream server when they have not found the requested content in the cache or when it is outdated. Modeling such services that can change their behavior based on the situation, although possible in \name, requires designing content-aware modules for the simulator. For this, interested users can extend the corresponding classes to override the methods responsible for consuming resources, as we well as the pattern of executing endpoint functions and their corresponding threads.

\subsubsection{Model portability}
We rigorously evaluated \name{} in various scenarios and demonstrated its accuracy in different settings. However, we find out that even though the extracted models can be used in different hosts and settings, they cannot be generalised/used  to simulate the performance in other microarchitecture settings. For example, a model extracted for an Intel CPU architecture cannot be used to accurately simulate the performance of the ARM-compatible version of the software in an ARM-based architecture (e.g., Apple's M-series chips - even with dynamic binary translation enabled).

\subsubsection{Simulating highly overloaded scenarios}
Two paramount features of \name{} are (1) its ability to simulate contention for various resources when multiple containers are packed in a host, and (2) simulating network congestion when microservices communicate over a network topology. However, when a process enters an highly overloaded state (i.e., when the rate of incoming requests is more than the processing capacity of serving processes 
), the request queue will grow without a bound, and in the meanwhile, the system aggressively throttles CPU and network resources. This, in turn, will increase the rate of context switches, CPU migrations, and cache misses. 

In large-scale systems, performance engineers generally define a rigorous set of constraints and resource management policies throughout a software's design and deployment lifecycle to prevent the system from entering such unpredictable situations \cite{10.5555/542306}. Hence, accurate simulation of such chaotic situations are extremely challenging. 

\subsubsection{Simulating containers on system virtual machines}

One of the widely used schemes for deploying container orchestrators (e.g., Kubernetes) on a cluster involves the use of hypervisor-based virtualization platforms, such as OpenStack or VMWare. Such deployment scheme becomes popular as system virtual machines (1) allows the use of various operating systems in one machine, (2) allows OS-level customization for each service, and (3) provide better data and resource isolation between neighbouring services. However, study shows that virtualization suffers from noticeable performance overhead due to the additional OS layer \cite{8284700, Shirinbab2019PerformanceIO}. 

One of the main intentions of designing \name{} is to simulate thread-level performance of services and to enable accurate prediction of endpoint functions performance when various threads compete over resources in a host. Thus, the additional complexity of OS-level threads in the aforementioned schemes, if not appropriately modelled, may affect the accuracy of simulation in scenarios where the virtualization overhead is considerably high.

\subsubsection{Challenges of performance modeling}

As described earlier, the process of performance modeling occurs on a PTE that mimics a PE to (1) obtain a complete isolation for services to collect accurate measurements for each endpoint function and (2) to prevent interference with the experience of real cloud users. However, preparing a PTE may impose both technical and financial challenges that needs to be addressed in the design of DevOps process flows. Automating the entire process of performance testing after each release may dramatically reduce the cost of modeling and dramatically increase the accuracy of simulations.

\section{Conclusion and future works}

In this work, we presented \name{} as a systematic method and simulation platform for modeling and simulating the performance of large-scale service chains in the context of cloud native computing. Using performance tracing and monitoring tools, \name{} allows performance modeling of various microservices and their corresponding service chains in cloud native orchestration platforms (such as Kubernetes that we used in this article) and enables the possibility to simulate the effectiveness of different resource management scenarios and placement policies. We evaluated the accuracy of \name{} in a set of prevalent scenarios and obtained {\raise.17ex\hbox{$\scriptstyle\mathtt{\sim}$}}81-99\% prediction accuracy as well as {\raise.17ex\hbox{$\scriptstyle\mathtt{\sim}$}}16-1200 times speed-up factor in comparison to testing on a real system (excluding the time needed for setting up the system and configuring the cluster). We evaluated the capability of \name{} in simulating large-scale service chains and showed that using a single-core of a laptop, it can achieve a 10 fold speed-up factor to simulate a scenario with complex service chains (composed of 100 microservices interconnected with 200 links deployed over a cluster with 100 hosts).

Our future work will cover three main categories: (1) improving speed, (2) improving accuracy, and (3) expanding use cases of \name. To increase its speed-up factor, we plan to refactor it to harness the parallel processing capabilities of GPUs. To improve its accuracy, we will add more layers/features, such as the support of defining cache-size per container (to simulate various scenarios based on Intel's CAT) or adding the simulation of poll mode in NICs (e.g., scenarios involving DPDK). Because advanced performance optimization methods can harness \name{'s} fast scenario assessment capability to quickly perform numerous policy trials and errors, we will work towards proposing novel performance optimization methods. We will also use \name{} to train deep reinforcement learning agents that can autonomously control different policies in the cluster. We will also implement accounting features in PerfSim that would allow detailed report on cost of deployment (e.g., when deploying on a public cloud) based on pricing/cost information provided within the scenario config files.

\section*{Acknowledgments}
Parts of this work has been supported by the Knowledge Foundation of Sweden (KKS) through the Project AIDA (20200067). The contribution of Auday Al-Dulaimy to this work has been  performed with the support of the KKS under the SACSys project.

\ifCLASSOPTIONcaptionsoff
	\newpage
\fi

\bibliographystyle{IEEEtran}
\bibliography{IEEEabrv,bib}

\begin{IEEEbiography}[{\includegraphics[width=1in,height=1.25in,clip,keepaspectratio]{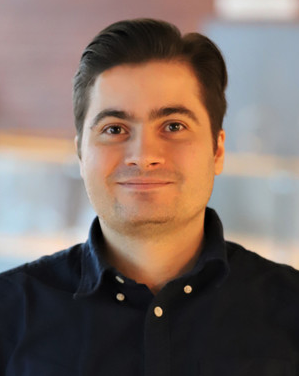}}]{Michel Gokan Khan} (M'13) received the M.S. degree in software engineering from Iran University of Science and Technology, Tehran, Iran, in 2016. He is currently pursuing the Ph.D. degree in computer science at Karlstad University, Sweden.
	His main research interest includes network function virtualization, cloud native applications, mathematical and machine learning based performance optimization for telecom clouds, cloud computing, and applications of AI in distributed systems.
\end{IEEEbiography}

\begin{IEEEbiography}[{\includegraphics[width=1in,height=1.25in,clip,keepaspectratio]{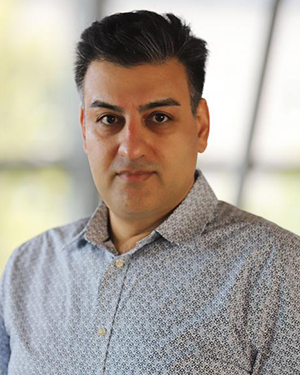}}]{Javid Taheri} received the Ph.D. degree in mobile computing from the School of Information Technologies, The University of Sydney, Australia. He is currently a Professor with the Department of Computer Science, Karlstad University, Sweden. His major areas of interest are profiling, modeling, and optimization techniques for private and public cloud infrastructures, profiling, modeling, and optimization techniques for software-defined networks, and network-aware scheduling algorithms for cloud 
computing.
\end{IEEEbiography}

\begin{IEEEbiography}
	[{\includegraphics[width=1in,height=1.25in,clip,keepaspectratio]{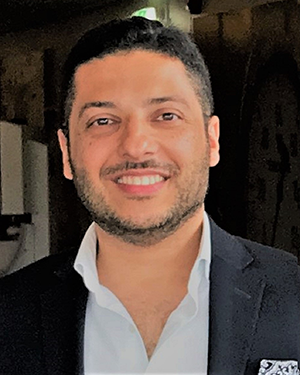}}]
	{Auday Al-Dulaimy}
	is a postdoctoral researcher at the School of Innovation, Design and Engineering, Mälardalen University, Sweden. He received PhD degree in Computer Science from Beirut Arab University, Lebanon in 2017; and B.Sc and M.Sc degrees in Computer Science from Al-Nahrain University, Iraq in 2000 and 2003 respectively. His major areas of research interest includes Distributed Systems, Cloud Computing, Internet of Things, and Edge Computing.
\end{IEEEbiography}

\begin{IEEEbiography}[{\includegraphics[width=1in,height=1.25in,clip,keepaspectratio]{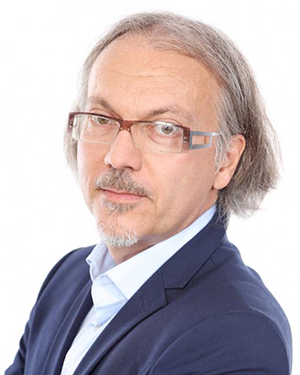}}]{Andreas J. Kassler} (SM'12) received his M.Sc. degree in mathematics/computer science from Augsburg University, Germany in 1995 and his Ph.D. degree in computer science from University of Ulm, Germany in 2002. He is currently Professor with the Department of Computer Science, Karlstad University, Sweden. He teaches wireless networks and advanced topics in computer networking. His main research interests include Software Defined Networks, Future Internet, and Network Function Virtualization.
\end{IEEEbiography}

\end{document}